\documentclass[article, nojss]{jss}\usepackage[]{graphicx}\usepackage[]{color}
\makeatletter
\def\maxwidth{ %
  \ifdim\Gin@nat@width>\linewidth
    \linewidth
  \else
    \Gin@nat@width
  \fi
}
\makeatother

\usepackage{Sweave}

\usepackage{thumbpdf,lmodern}

\usepackage{graphicx}
\usepackage{subcaption} 
\usepackage{algorithm}
\usepackage{multirow}
\usepackage{tikz}
\usetikzlibrary{positioning}
\usepackage{caption} 
\usepackage{amsmath}
\DeclareMathOperator*{\argmax}{arg\,max}

\author{Devin Incerti\\Genentech
   \And Jeroen P Jansen\\University of California, San Francisco}
\Plainauthor{Devin Incerti, Jeroen Jansen}

\title{\pkg{hesim}: Health Economic Simulation Modeling and Decision Analysis}
\Plaintitle{hesim: Health Economic Simulation Modeling and Decision Analysis}

\Abstract{
Health economic models simulate the costs and effects of health technologies for use in health technology assessment (HTA) to inform efficient use of scarce resources. Models have historically been developed using spreadsheet software (e.g., Microsoft \proglang{Excel}) and while use of \proglang{R} is growing, general purpose modeling software is still limited. \pkg{hesim} is an \proglang{R} package that helps fill this gap by facilitating parameterization, simulation, and analysis of economic models in an integrated manner. Supported model types include cohort discrete time state transition models (cDTSTMs), individual continuous time state transition models (iCTSTMs), and partitioned survival models (PSMs), encompassing Markov (time-homogeneous and time-inhomogeneous) and semi-Markov processes. A modular design based on \pkg{R6} and \code{S3} classes allows users to combine separate submodels for disease progression, costs, and utility in a flexible way. Probabilistic sensitivity analysis (PSA) is used to propagate uncertainty in model parameters to model outputs. Simulation code is written in \proglang{C++} so complex simulations such as those combining PSA and individual simulation can be run much more quickly than previously possible. Decision analysis within a cost-effectiveness framework is performed using simulated costs and quality-adjusted life years (QALYs) from a PSA.
}

\Keywords{health economic evaluation, cost-effectiveness analysis, simulation, multi-state models, \proglang{R}}
\Plainkeywords{health economic evaluation, cost-effectiveness analysis, simulation, multi-state models, R}

\Address{
  Devin Incerti\\
  Genentech\\
  South San Francisco\\
  E-mail: \email{incerti.devin@gene.com}\\
  
  Jeroen P Jansen\\
  University of California, San Francisco\\
  San Francisco\\
  E-mail: \email{jeroen.jansen@ucsf.edu}\\
}
\IfFileExists{upquote.sty}{\usepackage{upquote}}{}
\begin{document}



\section{Introduction} \label{sec:intro}
Health technology assessment (HTA) is a systematic approach for comparing competing health technologies to inform the efficient use of health care resources. Publicly funded health systems such as those in Australia, Canada, and the United Kingdom, among others, use HTA to help maximize health gains for the population given a fixed budget. HTA is also used by private payers to guide coverage decisions and the adoption of new technologies \citep{trosman2011health}. Such assessments usually rely heavily on cost-effectiveness analysis (CEA) so that decisions can be made on the basis of a formal evaluation of costs and effects \citep{dakin2015influence}. CEA is made feasible by development of health economic models that simulate costs and effects over relevant time horizons for the compared health technologies using the totality of relevant evidence for the decision problem at hand. 

The most commonly used economic models are Markov state transition models (STMs) that simulate transitions between mutually exclusive health states. When the Markov assumption holds so that transition probabilities are either constant over time or depend only on model time, a cohort-level model can be used \citep{briggs1998introduction}. If the Markov assumption is relaxed---e.g., with a semi-Markov model so that transition probabilities depend on time in an intermediate health state---then individual-level models are required in most cases \citep{brennan2006taxonomy, fiocco2008reduced}. Individual-level models afford considerably more flexibility and allow patient history to be tracked over time.

Decisions informed by economic models and CEA are subject to uncertainty. One source of decision uncertainty stems from uncertainty in the underlying model parameters. Parameter uncertainty is typically quantified using probabilistic sensitivity analysis (PSA), which involves randomly sampling the model parameters from suitable probability distributions and simulating the model for each sampled parameter set \citep{claxton2005probabilistic}. When combined with individual-level simulation, PSA can take an appreciable amount of time to run \citep{ohagan2007monte}.

Estimation of model parameters should ideally be performed using statistical models that are aligned with the structure of the economic model. For instance, when patient-level data is available, a multi-state model can be used to parameterize all possible transitions in a STM while accounting for censoring and competing risks \citep{williams2017cost}. Similarly, in oncology, partitioned survival models (PSMs) can be parameterized from estimates of progression-free survival (PFS) and overall survival (OS). In other cases, parameters might be combined from disparate sources, such as within a single Bayesian model \citep{baio2012bayesian}. When the clinical evidence base is not limited to a single study, a formal evidence synthesis, such as a network-meta analysis (NMA), might even be performed \citep{dias2018network}.

Despite their computational demands and foundations in statistics, health economic models have historically been developed with specialized commercial software (e.g., \proglang{TreeAge}) or more commonly with a spreadsheet (almost always Microsoft \proglang{Excel}). The limitations of such software relative to programming languages like \proglang{R} have been increasingly emphasized in the literature \citep{baio2017simple, incerti2019r, jalal2017overview}. It is therefore no surprise that a number of related \proglang{R} packages have recently been developed, such as \pkg{BCEA} \citep{baio2017bayesian}, \pkg{SAVI} \citep{strong2014estimating}, \pkg{dampack} \citep{alarid2018dampack}, \pkg{EVSI} \citep{heath2018efficient}, \pkg{survHE} \citep{baio2020survhe}, and \pkg{heemod} \citep{filipovic2017markov}. Still, of the available packages, only \pkg{heemod} provides a general purpose framework for developing simulation models and it is limited to Markov cohort models. 

\pkg{hesim} is an \proglang{R} package that advances the functionality and performance of the existing software. Multiple model types are supported including cohort discrete time state transition models (cDTSTMs), N-state PSMs, and individual-level continuous time state transition models (iCTSTMs), encompassing both Markov (time-homogeneous and time-inhomogeneous) and semi-Markov processes. To maximize flexibility and facilitate integration of the statistical methods and economic model, parameters can be set by either fitting models in \proglang{R} or by inputting values obtained from external sources. So that individual-level simulation and PSA can be run quickly, \pkg{Rcpp} and \pkg{data.table} are heavily utilized. After simulating costs and quality-adjusted life-years (QALYs) from a PSA, decision analysis can be performed within a cost-effectiveness framework.

The remainder of this paper is organized as follows. Section~\ref{sec:model-taxonomy} provides a mathematical description of the economic models supported by \pkg{hesim}. An overview of the coding framework is provided in Section~\ref{sec:framework}. An illustrative example using a three state model of disease progression in oncology is provided in Section~\ref{sec:illustrative-example}. Analyses are conducted for each of the three supported model types, illustrating approaches suitable for both patient-level and aggregate-level data. The cost-effectiveness framework is described in Section~\ref{sec:cea} along with worked examples based on the output from the prior section. Section~\ref{sec:discussion} discusses additional features not covered in this paper, makes comparisons to other software, and explores possible extensions. Finally, Section~\ref{sec:conclusion} concludes. 

Analyses were performed using \pkg{hesim} version \code{0.5.0}. All code including a replication \proglang{R} script for this paper is available at \url{https://github.com/hesim-dev/hesim-manuscript}.

\section{Model taxonomy} \label{sec:model-taxonomy}
STMs simulate transitions between mutually exclusive health states. A common assumption is that a STM is a Markov model, meaning that transitions to the next health state can only depend on the present health state. In a time homogeneous Markov model transition probabilities are constant over time, whereas in a time inhomogeneous model they can depend on time since the start of the model. A semi-Markov model relaxes the Markov assumption and allows transitions to depend on time since entering an intermediate state.

Markov and semi-Markov models can be formulated in either continuous or discrete time, at either the cohort- or individual-level. In health economics, Markov cohort models tend to be formulated in discrete time (i.e., as cDTSTMs), although state probabilities can be computed with continuous time models using the Aalen-Johansen estimator \citep{aalen1978empirical} or the Kolmogorov forward equation \citep{cox1977theory}. While tunnel states can be used to approximate a semi-Markov model using a cohort approach, they can only be simulated in a general fashion using individual-level models. Discrete time individual simulation is possible, but we use continuous time models (i.e., iCTSTMs) because they do not require specification of model cycles and can be run considerably faster. 

PSMs are specialized models that can be parameterized using survival curves and are especially useful in oncology where PFS and OS are commonly reported. They are ``area under the curve'' models, although they can also be formulated as STMs by using the survival curves to construct transition probabilities. 

\subsection{Cohort discrete time state transition models} \label{sec:cDTSTMs}
cDTSTMs simulate the probability that a cohort of patients is in each of $H$ health states over time. Time is partitioned into intervals $\{[0, t_1], [t_1, t_2], \ldots, [t_{C-1}, t_C]\}$ with each interval $1,\ldots, C$ referred to as a model cycle. We denote $t_c$ as the time at the end of model cycle $c$. An $H \times 1$ state vector that stores the probability of being in each health state at cycle $c$ is written as $x_c = (x_{1c}, x_{2c}, \ldots, x_{Hc})$ where $\sum_h x_{hc} = 1$. A transition probability matrix is an $H \times H$ matrix denoted by $P_c$ where the $(r,s)$th element represents a transition from state $r$ to state $s$ between model cycles $c$ and $c+1$. The state vector at cycle $c+1$ for each of $C$ model cycles is given by, 

\begin{equation} \label{eqn:markov-sim}
x_{c+1}^T= x_c^T P_c, \quad c = 0,\ldots, C-1.
\end{equation}

The transition probability matrices can also be characterized in terms of transition intensity matrices, $Q_c$, of the same dimensions as $P_c$ and representing an underlying continuous process with a patient in state $X(t)$ at exact time $t$. The $(r,s)$th element is the hazard representing the instantaneous risk of moving from state $r$ to state $s$,

\begin{equation} \label{eqn:hazard}
h_{rs}(t) = \lim_{\Delta t\to 0} \frac{\Prob(X(t + \Delta t) = s | X(t) = r)}{\Delta t}.
\end{equation}

Each row of $Q_c$ sums to 0 whereas each row of $P_c$ sums to 1. If transition intensities are constant over the interval $(t_c, t_{c+1})$ for a model cycle with length $u = t_{c+1}-t_c$, then the transition probability matrix can be solved with the matrix exponential as,

\begin{equation} \label{eqn:expmat}
P_c = \textrm{Exp}(uQ_c).
\end{equation}

Costs and QALYs are computed by assigning (potentially time-varying) values to each health state. Utility, a measure of preference for a health state that normally ranges from $0$ (dead) to $1$ (perfect health), is used when computing QALYs \citep{torrance1986measurement}. Assuming model time is in years, state values for costs are estimated by annualizing costs. In a Markov model, state values, like transition probabilities, can depend on time since the start of the model but not on time since entering an intermediate health state.

Expected values are computed by integrating the probability of being in each state and ``weighted'' state values, where weights are a function of the discount factor. That is, for a time horizon $T$, discounted costs and QALYs in health state h are computed as,

\begin{equation} \label{eqn:cohort-costs-qalys}
\int_0^{T} z_h(t) e^{-rt} p_h(t) dt,
\end{equation}

where $r$ is the discount rate and at time $t$, $z_h(t)$ is the predicted cost or utility value and $p_{h}(t) = \Prob(X(t) = h)$ is the probability of being in health state $h$. 

In discrete time, the integral is approximated with

\begin{equation} \label{eqn:riemann-sum}
\sum_{c = 1}^{C}f(t_c^{*})\Delta t_c,
\end{equation}

where $\Delta t_c = t_{c} - t_{c-1}$,  $t_c^{*}\in[t_{c-1},t_{c}]$, and $f(t)= z_h(t) e^{-rt} p_h(t)$. Three methods can be used to estimate $f(t_c^{*})$. First, a left Riemann sum uses values at the start of each time interval, $f(t_c^{*}) = f(t_{c-1})$. Second, a right Riemann sum uses values at the end of each time interval, $f(t_c^{*}) = f(t_{c})$. Finally, the trapezoid rule averages values at the start and end of each interval, $f(t_c^{*}) = \frac{1}{2}\Delta t_c[f(t_{c-1}) + f(t_{c})]$. 

\subsection{Individual continuous time state transition models} \label{sec:iCTSTMs}
iCTSTMs simulate individual trajectories between health states using random number generation. Trajectories are simulated for multiple patients and expected costs and QALYs are computed by averaging across the simulated patients. A reasonably large number of patients must be simulated to ensure that expected values are stable \citep{ohagan2007monte}. 

In continuous time, a patient is in state $X(t)$ at time $t$. State transitions are modeled using a multi-state modeling framework \citep{putter2007tutorial} where the probability of a transition from state $r$ to state $s$ is governed by the hazard function from Equation~\ref{eqn:hazard}.

Simulated disease progression is characterized by $J$ distinct jumps between health states $D = \{(t_0, X(t_0)), (t_1, X(t_1)), \ldots (t_J, X(t_J))\}$ with a patient remaining in a health state from time $t_j$ until transitioning to the next state at $t_{j+1}$. Jumps between health states are simulated using parametric and flexible parametric survival models as implemented in the \pkg{flexsurv} package \citep{jackson2016flexsurv}. Specifically, if a patient enters state $r$ at time $t_j$, then a probability density function for the time-to-event $t^*$ for the $r \rightarrow s$ transition is, 

\begin{equation}
f_{rs}(t^*|\theta(z), t_j), \quad t^*\geq 0,
\end{equation}

where parameters $\theta = (\theta_1, \theta_2, \ldots, \theta_p)$ may depend on covariates $z_p$ through the link function $g(\theta_p) = z_p^T \gamma$ and $\gamma$ is a vector of regression coefficients. In a time inhomogeneous Markov model, time $t^*=t_{j+1}$ is conditional on not experiencing event $s$ until time $t_j$ (i.e., it is left-truncated at time $t_j$). In a semi-Markov model, time-to-event is expressed in terms of $t^* = t_{j+1}-t_j$ and a patient enters state $s$ at time $t_j + t^*$.

A survival distribution is specified for each permitted transition in the multi-state model. A trajectory through the model can then be simulated as described in Algorithm~\ref{alg:sim-ictstm}. While the algorithm repeats until a patient dies, it can also be stopped at a specified time $t$ or when a patient reaches a maximum age. In the latter scenario, death is assumed to occur at the maximum age.
 
\begin{algorithm}
\caption{Simulation of individual continuous time state transition model.}
\label{alg:sim-ictstm}
\begin{enumerate}
\item Let $r$ be the state entered at time $t_j$. The number of permitted transitions from state $r$ is given by $n_r$. If $j=0$, then $t_j = 0$. 
\item Simulate times $\mathcal{T} = \{t_{1, j+1}, t_{2, j+1}, \ldots t_{n_r, j+1}\}$ to each of the $n_r$ permitted transitions.
\item Set the time of the transition $t_{j+1}$ equal to the minimum simulated time in $\mathcal{T}$ and the next state $s$ to the state with the minimum simulated time.
\item Set $r=s$ and $t_j = t_{j+1}$. If the patient is still alive, repeat the previous steps until death.
\end{enumerate}
\end{algorithm}

Costs and QALYs are computed using the continuous time present value given a flow of state values, which change as patients transition between health states or as costs vary as a function of time. The state values can be partitioned into $M$ time intervals indexed by $m = 1,\ldots, M$ where interval $m$ contains times $t$ such that $t_m < t \leq t_{m+1}$ and values for state $h$ are equal to $z_{hm}$ during interval $m$. $z_{hm}$ will equal zero during time intervals in which a patient is not in state $h$. Discounted costs and QALYs for health state $h$ are then given by,  

\begin{equation}
\sum_{m = 1}^M \int_{t_m}^{t_m+1} z_{hm}e^{-rt}dt = \sum_{m = 1}^M z_{hm} \left(\frac{e^{-r{t_{m}}} - e^{-r{t_{m+1}}}}{r}\right),
\end{equation}

where $r > 0$ is again the discount rate. If $r = 0$, then the present value simplifies to $\sum_{m = 1}^M z_{hm}(t_{m+1} - t_{m})$. 

Note that while state values in cohort models can depend on time since the start of the model, state values in individual-level models can depend on either time since the start of the model or time since entering the most recent health state. Individual-level models consequently not only afford more flexibility than cohort models when simulating disease progression, but when simulating costs and/or QALYs as well.

\subsection{Partitioned survival models} \label{sec:PSMs}
PSMs are conceptually similar to STMs in that they are characterized by mutually exclusive health states. They differ, however, in that state probabilities are not computed via matrix multiplication or individual simulation, but from a set of non-mutually exclusive survival curves \citep{glasziou1990quality, woods2017nice}. Each survival curve represents time to transition to that state or to a more severe health state.

In an $N$-state model, $N-1$ non-mutually exclusive survival curves are required. The cumulative survival function, $S_n(t)$, represents the probability that a patient survives to health state $n$ or to a lower indexed state beyond time $t$. The probability that a patient is in health state 1 is $S_1(t)$. State membership in health states $2,\ldots, N-1$ is computed as $S_{n}(t) - S_{n-1}(t)$. Finally, the probability of being in the final health state (i.e., the death state) is $1-S_{N-1}(t)$, or one minus overall survival function.

Survival functions are estimated by fitting (flexible) parametric survival models as described in Section~\ref{sec:iCTSTMs}. The $n$th fitted survival model with density $f_n(t)$ has cumulative density function $F_n(t)$, survivor function $1 - F_n(T)$, cumulative hazard $H_n(t) = -\log S_n(t)$, and hazard $h_n(t) = f_n(t)/S_n(t)$. State probabilities for each health state, $x_{hc}$, can be computed for an arbitrarily fine grid of time periods $c$. Costs and QALYs are then computed in the same manner as in the cohort model described in Section~\ref{sec:cDTSTMs}.

\section{Framework} \label{sec:framework}
Economic models consist of a disease model, a utility model, and a set of cost models for each cost category. Model development proceeds in 4 steps. The first step is to set up the model by defining the model structure, target population, and treatment strategies of interest. The relevant information is then stored in a \code{hesim_data} object. 

\begin{figure}[h]
\centering
\begin{tikzpicture}
\node[draw, minimum height=4cm, minimum width=3.5cm, font=\footnotesize, text depth = 3.5cm] (A) at (-5,0) {\textbf{1. Parameterization}};
\node[draw,fill=white, minimum height=.8cm, minimum width=2.5cm, font=\footnotesize] at (-5, .8){Disease model};
\node[draw,fill=white, minimum height=.8cm, minimum width=2.5cm, font=\footnotesize] at (-5, -.2){Utility model};
\node[draw,fill=white, minimum height=.8cm, minimum width=2.5cm, font=\footnotesize] at (-5, -1.2){Cost model(s)};

\node[draw, minimum height=4cm, minimum width=3.5cm, font=\footnotesize, text depth = 3.5cm] (B) at (0,0) {\textbf{2. Simulation}};
\node[draw,fill=white, minimum height=.8cm, minimum width=2.5cm, font=\footnotesize] at (0, .8){Disease model};
\node[draw,fill=white, minimum height=.8cm, minimum width=2.5cm, font=\footnotesize] at (0, -.2){Utility model};
\node[draw,fill=white, minimum height=.8cm, minimum width=2.5cm, font=\footnotesize] at (0, -1.2){Cost model(s)};

\node[draw, minimum height=4cm, minimum width=3.5cm, font=\footnotesize, text depth = 3.5cm] (C) at (5,0) {\textbf{3. Decision analysis}};
\node[draw,fill=white, minimum height=.8cm, minimum width=2.5cm, font=\footnotesize] at (5, .3){CEA};
\node[draw,fill=white, minimum height=.8cm, minimum width=2.5cm, font=\footnotesize] at (5, -.7){MCDA};

\node[draw, above = .5 of A, minimum width=3cm, font=\footnotesize] (D) {Estimation data};
\node[draw, above = .5 of B, minimum width=3cm, font=\footnotesize] (E) {Input data};

\draw [thick, ->] (A.east) -- (B.west);
\draw [thick, ->] (B.east) -- (C.west);
\draw [thick, ->] (D.south) -- (A.north);
\draw [thick, ->] (E.south) -- (B.north);

\end{tikzpicture}
\caption{Economic modeling process.}\label{fig:econ-eval-process-hesim}
\end{figure}
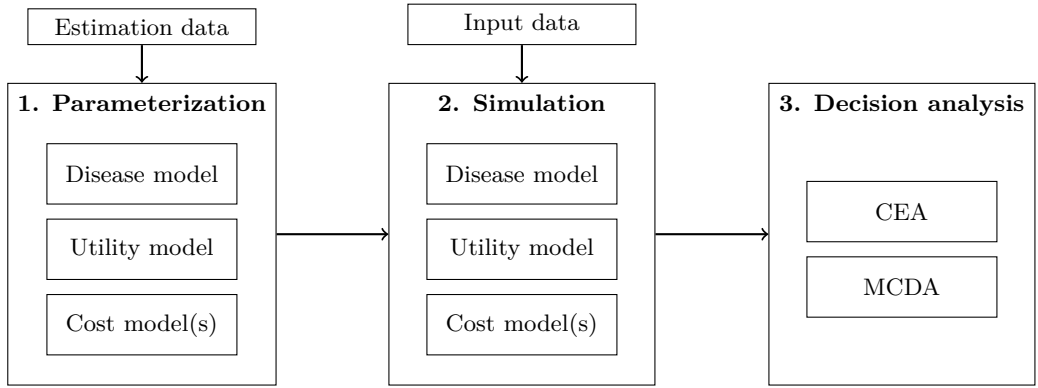

Development and analysis of the model is performed in the next three steps as shown in Figure~\ref{fig:econ-eval-process-hesim}. First, the disease progression, costs, and utility submodels are parameterized using ``estimation'' datasets. Next, the submodels are combined to construct an economic model. The economic model is then used to simulate disease progression, QALYs, and costs as a function of ``input data''. The input data always contains variables describing the target population and treatment strategies of interest, but can also contain variables related to time to facilitate use of time-varying covariates or parameters. Finally, the simulated QALYs and costs are used to perform decision analysis within a cost-effectiveness framework. While other approaches such as multi-criteria decision analysis (MCDA) could, in principle, be used as well, only CEA is currently supported. All models are simulated using PSA so that decision uncertainty can be represented.  

The three types of economic models described in Section~\ref{sec:model-taxonomy} are implemented as the \pkg{R6} classes \code{CohortDtstm}, \code{IndivCtstm}, and \code{Psm}. Each class contains methods for simulating disease progression, QALYs, and costs. These methods are made possible by separate \pkg{R6} classes for the disease, utility, and cost submodels. The remainder of this section describes the framework for parameterization and simulation. The cost-effectiveness framework is described in Section~\ref{sec:cea}. 

\subsection{Parameterization} \label{sec:parameterization}
Each submodel contains fields for the model parameters and the input data. Parameters can either be estimated from a statistical model fit using \proglang{R} or input directly by the user. There are two types of parameter objects, standard parameter objects prefixed by \code{``params''} and ``transformed'' parameter objects prefixed by \code{``tparams''}. The former contain the underlying parameters of a statistical model while the latter contain parameters more immediate to prediction that have been transformed as a function of the input data. The regression coefficients of a logistic regression are an example of parameter objects while the predicted probabilities are examples of a transformed parameter object.

Transformed parameter objects are often useful when parameterizing a model using external sources. Two common examples are the predicted probabilities comprising transition probability matrices and predicted means for assigning values to health states. The advantage of a transformed parameter object is that an explicit statistical model does not need to be specified which gives the user more flexibility.

\subsubsection{Disease progression}
The statistical model used to parameterize the disease model depends on the type of economic model. For example, multinomial logistic regressions can be used to parameterize a cDTSTM, a set of N-1 independent survival models are used to parameterize an N-state partitioned survival model, and multi-state models can be used to parameterize an iCTSTM (Table~\ref{tbl:parameterize-disease-model}). 

\begin{table} [h]
\caption{Parameterization of disease models.}\label{tbl:parameterize-disease-model}
\footnotesize
\begin{tabular*}{\textwidth}{l l l l}
\hline
Economic model & Statistical model & Parameter object & Model object\\
\hline
\code{CohortDtstm} & Custom & \code{tparams_transprobs} & \code{msm::msm} \\
 & Multinomial logistic regressions & \code{params_mlogit} & \code{multinom_list}\\
\code{IndivCtstm} & Multi-state model (joint likelihood) & \code{params_surv} & \code{flexsurv::flexsurvreg}\\
& Multi-state model (transition-specific) & \code{params_surv_list} & \code{flexsurvreg_list}\\
\code{Psm} & Independent survival models & \code{param_surv_list} & \code{flexsurvreg_list}\\
\hline
\end{tabular*}
\end{table}

The parameters of a survival model are stored in a \code{params_surv} object and a \code{params_surv_list} can be used to store the parameters of multiple survival models. The latter is useful for storing the parameters of a multi-state model or the independent survival models required for a PSM. The parameters of a multinomial logistic regression are stored in a \code{params_mlogit} object and can be created by fitting a model for each row in a transition probability matrix with \code{nnet::multinom()}.

\code{tparams_transprobs} objects store transition probability matrices that have been predicted for each PSA sample, treatment strategy, patient from the target population, and optionally time interval (to allow for time inhomogeneous Markov models). Transition probabilities can be predicted from a fitted multi-state model using the \pkg{msm} package or constructed ``by hand'' in a custom manner. 

\subsubsection{Costs and utility} 
State values (i.e., costs and utilities) do not depend on the choice of disease model. They can currently either be modeled using a linear model or using predicted means (Table~\ref{tbl:parameterize-stateval-model}). 

\begin{table} [h]
\caption{Parameterization of state value models.}\label{tbl:parameterize-stateval-model}
\footnotesize
\begin{tabular*}{\textwidth}{@{\extracolsep{\fill}}l l l l}
\hline
Statistical model & Parameter object & Model object\\
\hline
Predicted means & \code{tparams_mean} & \code{stateval_tbl}\\
Linear model & \code{params_lm} & \code{stats::lm}\\
\hline
\end{tabular*}
\end{table}

The most straightforward way to construct state values is with \code{stateval_tbl}, which is a special object used to assign values (i.e., predicted means) to health states that can vary across PSA samples, treatment strategies, patients, and/or time intervals. State values can be specified either as moments (e.g., mean and standard error) or parameters (e.g., shape and scale of gamma distribution) of a probability distribution, or by pre-simulating values from a suitable probability distribution (e.g., from a Bayesian model).

\subsection{Simulation} \label{sec:simulation}
The utility and cost models are always \pkg{R6} objects of class \code{StateVals}, whereas the disease models vary by economic model (Table~\ref{tbl:submodels}). The disease model is used to simulate health state transitions in a cDTSTM and iCTSTM, and survival curves in a PSM.

\begin{table} [h]
\caption{Submodels comprising an economic model.}\label{tbl:submodels}
\footnotesize
\begin{tabular*}{\textwidth}{@{\extracolsep{\fill}}l l l l}
\hline
Economic model & Disease model & Utility model & Cost model(s)\\
\hline
\code{CohortDtstm} & \code{CohortDtstmTrans} & \code{StateVals} & \code{StateVals}\\
\code{IndivCtstm} & \code{IndivCtstmTrans} & \code{StateVals} & \code{StateVals}\\
\code{Psm} & \code{PsmCurves} & \code{StateVals} & \code{StateVals}\\
\hline
\end{tabular*}
\end{table}

The disease and state value models can either be instantiated from parameter or model objects using \code{S3} generic functions prefixed by ``\code{create}'' or using the \pkg{R6} constructor method \code{$new()}. \code{create_IndivCtstmTrans()} is an example of the former that can be used to create an \code{IndivCtstmTrans} object from a \code{flexsurvreg}, \code{flexsurvreg_list}, \code{params_surv}, or \code{params_surv_list} object. Similarly, a \code{StateVals} object can, for example, be created from a \code{stateval_tbl} object with \code{create_StateVals()}. The complete economic model is instantiated by combining the submodels using \code{$new()}.

Each economic model contains methods for simulating disease progression, QALYs, and costs (Table~\ref{tbl:simulate-outcomes}). The utility and cost models always simulate QALYs and costs from the simulated progression of disease with the methods \code{$sim_qalys()} and \code{$sim_costs()}, respectively. Methods for simulating disease progression differ slightly since the simulation approaches differ as described in Section~\ref{sec:model-taxonomy}. In an N-state PSM $N-1$ survival curves are generated and in an iCTSTM a trajectory through the STM is simulated for multiple patients via individual simulation. All models can simulate state probabilities over time.

\begin{table} [h]
\caption{Methods for simulating outcomes from economic models.}\label{tbl:simulate-outcomes}
\footnotesize
\begin{tabular*}{\textwidth}{@{\extracolsep{\fill}}l l l l}
\hline
Economic model & Disease progression  & QALYs & Costs \\
\hline
\code{CohortDtstm} & \code{sim_stateprobs()} & \code{sim_qalys()} & \code{sim_costs()}\\
\code{IndivCtstm} & \code{sim_disease()}, \code{sim_stateprobs()}  & \code{sim_qalys()} & \code{sim_costs()}\\
\code{Psm} & \code{sim_survival()}, \code{sim_stateprobs()}  & \code{sim_qalys()} & \code{sim_costs()}\\
\hline
\end{tabular*}
\end{table}

\section{Illustrative example} \label{sec:illustrative-example}
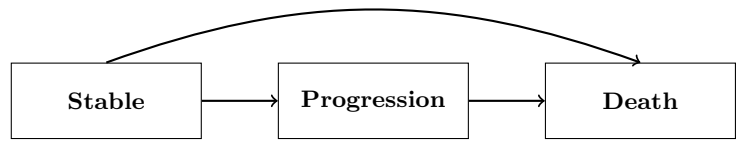
\begin{figure}[h]
\centering
\begin{tikzpicture}
\node[draw, minimum height=1cm, minimum width=2.5cm, font=\footnotesize] (S) {\textbf{Stable}};
\node[draw, minimum height=1cm, minimum width=2.5cm, font=\footnotesize] (P) [right = of S] {\textbf{Progression}};
\node[draw, minimum height=1cm, minimum width=2.5cm, font=\footnotesize] (D) [right = of P] {\textbf{Death}};

\draw [thick, ->] (S.east) -- (P.west);
\draw [thick, ->] (P.east) -- (D.west);
\draw [thick, ->] (S.north) to [bend left = 20] (D.north);

\end{tikzpicture}
\caption{Three state model of disease progression in oncology.}\label{fig:three-state-model}
\end{figure}

We consider a three-state model commonly used in oncology. As shown in Figure~\ref{fig:three-state-model}, the three health states are stable disease (i.e., not progressed), progression, and death. Patients can transition to a more severe health state (stable $\rightarrow$ progression, stable $\rightarrow$ death, and progression $\rightarrow$ death) but cannot recover to a less severe health state. We assume that patients remain on first line (1L) treatment until progression, at which time they switch to a second line (2L) treatment.

This model can be simulated using any of the model types described in Section~\ref{sec:model-taxonomy}. We demonstrate use of all three below, noting that the type of model employed should depend on the available data. We begin with the richest dataset in Section~\ref{sec:ictstm-example} and show how a flexible iCTSTM can be parameterized when continuously observed patient-level multi-state data is available. The next two sections then show how models can be developed when the data is more limited. Section~\ref{sec:psm-example} uses a PSM in a case where PFS and OS are available but multi-state data is not, and Section~\ref{sec:cdtstm-example} develops a cDTSTM in a setting where multi-state data is available but it is not continuously observed. For the sake of illustration, in both Sections~\ref{sec:psm-example} and~\ref{sec:cdtstm-example}, patient-level data is available for the SOC, but not for estimation of relative treatment effects.

\subsection{Individual continuous time state transition model} \label{sec:ictstm-example}
We set up the model by defining the model structure, target population, and treatment strategies of interest. The transitions are characterized by a matrix where each element denotes a transition from a row to a column. If a transition is not possible it is marked with \code{NA}; otherwise, an integer denotes the transition number.

\begin{Schunk}
\begin{Sinput}
R> tmat <- rbind(
+    c(NA, 1, 2),
+    c(NA, NA, 3),
+    c(NA, NA, NA)
+  )
R> colnames(tmat) <- rownames(tmat) <- c("Stable", "Progression", "Dead")
R> print(tmat)
\end{Sinput}
\begin{Soutput}
            Stable Progression Dead
Stable          NA           1    2
Progression     NA          NA    3
Dead            NA          NA   NA
\end{Soutput}
\end{Schunk}

The target population, treatment strategies, and (non-death) health states are stored in a \code{hesim_data} object and identified with integer valued identification variables (\code{patient_id}, \code{strategy_id}, and \code{state_id}). There are 3 treatment strategies: standard of care (SOC) and two new interventions (New 1 and New 2). The target population consists of a heterogeneous population of 1,000 patients, which is large enough to ensure that averages across patients in the individual simulation are reasonably stable. As we will see in the next section, patients can also be placed into subgroups (with a \code{grp_id} column) or given weights (with a \code{patient_wt} column) in the \code{patients} table; if left unspecified as done here, then all patients are assumed to belong to a single subgroup and are weighted equally.

\begin{Schunk}
\begin{Sinput}
> library("hesim")
> library("data.table")
> n_patients <- 1000
> patients <- data.table(
+   patient_id = 1:n_patients,
+   age = rnorm(n_patients, mean = 45, sd = 7),
+   female = rbinom(n_patients, size = 1, prob = .51)
+ )
> 
> states <- data.table(
+   state_id = c(1, 2),
+   state_name = c("Stable", "Progression") 
+ )
> 
> strategies <- data.table(
+   strategy_id = 1:3,
+   strategy_name = c("SOC", "New 1", "New 2"),
+   soc = c(1, 0, 0),
+   new1 = c(0, 1, 0),
+   new2 = c(0, 0, 1)
+ )
> 
> hesim_dat <- hesim_data(
+   strategies = strategies,
+   patients = patients,
+   states = states
+ )
> print(hesim_dat)
\end{Sinput}
\begin{Soutput}
$strategies
   strategy_id strategy_name soc new1 new2
1:           1           SOC   1    0    0
2:           2         New 1   0    1    0
3:           3         New 2   0    0    1

$patients
      patient_id      age female
   1:          1 46.26366      1
   2:          2 50.49314      1
   3:          3 35.52785      1
   4:          4 58.88309      0
   5:          5 53.66930      0
  ---                           
 996:        996 54.60555      0
 997:        997 50.10654      0
 998:        998 45.02035      0
 999:        999 45.60573      1
1000:       1000 39.45426      0

$states
   state_id  state_name
1:        1      Stable
2:        2 Progression

attr(,"class")
[1] "hesim_data"
\end{Soutput}
\end{Schunk}

Labels for the identification variables in a \code{hesim_data} object can be generated with \code{get_labels()}. These can (as will be shown below), in turn, be passed to plotting and summary functions to create tables and figures with more informative labels. 

\begin{Schunk}
\begin{Sinput}
R> labs_indiv <- get_labels(hesim_dat)
R> print(labs_indiv)
\end{Sinput}
\begin{Soutput}
$strategy_id
  SOC New 1 New 2 
    1     2     3 

$state_id
     Stable Progression       Death 
          1           2           3 
\end{Soutput}
\end{Schunk}

\subsubsection{Parameterization}
In a STM, the disease model governs transitions between health states. For now, we assume that individual patient data (IPD) with exact transition times is available so that a multi-state statistical model can be used to parameterize the disease model. 

Multi-state data consists of one row for each possible transition from a given health state where at most one transition is observed and all others are censored. We use the simulated dataset \code{onc3} from the \pkg{hesim} package.

\begin{Schunk}
\begin{Sinput}
R> onc3[patient_id %in% c(1, 2)]
\end{Sinput}
\begin{Soutput}
          from          to strategy_name female      age patient_id
1:      Stable Progression         New 2      0 59.85813          1
2:      Stable       Death         New 2      0 59.85813          1
3: Progression       Death         New 2      0 59.85813          1
4:      Stable Progression         New 2      0 62.57282          2
5:      Stable       Death         New 2      0 62.57282          2
   time_start time_stop status transition_id strategy_id      time
1:   0.000000  2.420226      1             1           3  2.420226
2:   0.000000  2.420226      0             2           3  2.420226
3:   2.420226 14.620258      1             3           3 12.200032
4:   0.000000  7.497464      0             1           3  7.497464
5:   0.000000  7.497464      0             2           3  7.497464
\end{Soutput}
\end{Schunk}

Multi-state models can be fit in one of two ways. First, a joint survival model with interaction terms for different transitions can be estimated, which is useful if there are constraints in the parameters, such as coefficients that are assumed equal across transitions. Here, we will take a second approach that is computationally more efficient and fit a separate model for each transition. We illustrate with parametric Weibull models but multiple distributions should be compared and evaluated in a real application \citep{williams2017cost}. The treatment effect is assumed to only influence transitions from stable disease since patients are assumed to switch to a 2L treatment after progression.

\begin{Schunk}
\begin{Sinput}
R> library("flexsurv")
R> n_trans <- max(tmat, na.rm = TRUE) 
R> wei_fits <- vector(length = n_trans, mode = "list")
R> f <- as.formula(Surv(time, status) ~ factor(strategy_name) + female + age)
R> for (i in 1:length(wei_fits)){
+    if (i == 3) f <- update(f, .~.-factor(strategy_name)) 
+    wei_fits[[i]] <- flexsurvreg(f, data = onc3, 
+                                 subset = (transition_id == i),
+                                 dist = "weibull")
+  }
R> wei_fits <- flexsurvreg_list(wei_fits)
\end{Sinput}
\end{Schunk}

Utility and cost values are stored in \code{stateval_tbl} objects. Utility is assumed to equal 0.8 with stable disease (standard error (SE) = 0.02) and 0.6 (SE = 0.05) with progressed disease. Utility values are assumed to follow a beta distribution, which can be parameterized either using its shape parameters or via the mean and standard error (in which case the method of moments is used to derive the shape parameters).

\begin{Schunk}
\begin{Sinput}
R> utility_tbl <- stateval_tbl(
+    data.table(state_id = states$state_id,
+               mean = c(.8, .6),
+               se = c(0.02, .05)
+    ),
+    dist = "beta")
R> print(utility_tbl)
\end{Sinput}
\begin{Soutput}
   state_id mean   se
1:        1  0.8 0.02
2:        2  0.6 0.05
\end{Soutput}
\end{Schunk}

Both medical and drug costs are considered. Medical costs are assumed to follow a gamma distribution which is often appropriate for costs since they tend to be right skewed. Like utility, the distribution is characterized by the mean (\$2,000 with stable disease, \$9,500 with progressed disease) and SE (assumed equal to the mean) and the method of moments is used to derive the underlying shape and scale parameters.

\begin{Schunk}
\begin{Sinput}
R> medcost_tbl <- stateval_tbl(
+    data.table(state_id = states$state_id,
+               mean = c(2000, 9500),
+               se = c(2000, 9500)
+    ),
+    dist = "gamma")
R> print(medcost_tbl)
\end{Sinput}
\begin{Soutput}
   state_id mean   se
1:        1 2000 2000
2:        2 9500 9500
\end{Soutput}
\end{Schunk}

Drug costs are assumed to be fixed (i.e., measured without uncertainty). When on 1L treatment, costs are \$2,000 with SOC, \$12,000 with New 1, and \$15,000 with New 2. All patients are assumed to switch to the same treatment at 2L after progression, which costs \$1,500 for the first 3 months and \$1,200 thereafter.

\begin{Schunk}
\begin{Sinput}
R> drugcost_tbl <- stateval_tbl(
+    drugcost_dt,
+    dist = "fixed")
R> print(drugcost_tbl)
\end{Sinput}
\begin{Soutput}
    strategy_id state_id time_id time_start time_stop   est
 1:           1        1       1       0.00      0.25  2000
 2:           1        1       2       0.25       Inf  2000
 3:           1        2       1       0.00      0.25  1500
 4:           1        2       2       0.25       Inf  1200
 5:           2        1       1       0.00      0.25 12000
 6:           2        1       2       0.25       Inf 12000
 7:           2        2       1       0.00      0.25  1500
 8:           2        2       2       0.25       Inf  1200
 9:           3        1       1       0.00      0.25 15000
10:           3        1       2       0.25       Inf 15000
11:           3        2       1       0.00      0.25  1500
12:           3        2       2       0.25       Inf  1200
\end{Soutput}
\end{Schunk}

\subsubsection{Simulation}
Before constructing each submodel, we specify the number of parameter samples that will be drawn for the PSA.

\begin{Schunk}
\begin{Sinput}
R> n_samples <- 1000
\end{Sinput}
\end{Schunk}

The input data for the transition model consists of one row for each treatment strategy and patient and can be easily generated from a \code{hesim_data} object with the \code{expand()} function.

\begin{Schunk}
\begin{Sinput}
R> transmod_data <- expand(hesim_dat,
+                          by = c("strategies", "patients"))
R> head(transmod_data)
\end{Sinput}
\begin{Soutput}
   strategy_id patient_id strategy_name soc new1 new2      age female
1:           1          1           SOC   1    0    0 46.26366      1
2:           1          2           SOC   1    0    0 50.49314      1
3:           1          3           SOC   1    0    0 35.52785      1
4:           1          4           SOC   1    0    0 58.88309      0
5:           1          5           SOC   1    0    0 53.66930      0
6:           1          6           SOC   1    0    0 53.40432      1
\end{Soutput}
\end{Schunk}

The transition model is then constructed as a function of the parameters (from the fitted Weibull models) and the input data. We must also specify the allowed transitions, the desired number of PSA samples, the "clock" (recall that we used a clock reset approach when fitting the Weibull models), and the starting age of the patients. Parameters for the PSA are, by default, drawn from the multivariate normal distribution of the maximum likelihood estimate of the regression coefficients, although we make this explicit with the \code{uncertainty} argument. The starting age of the patients is useful because it allows us to specify a maximum age for patients during the simulation, ensuring that the simulated patients do not live unrealistically long lives. 

\begin{Schunk}
\begin{Sinput}
R> transmod <- create_IndivCtstmTrans(wei_fits, transmod_data,
+                                     trans_mat = tmat, n = n_samples,
+                                     uncertainty = "normal",
+                                     clock = "reset",
+                                     start_age = patients$age)
\end{Sinput}
\end{Schunk}

The utility and cost models can be easily constructed from the \code{stateval_tbl} objects with the \code{create_StateVals()} function, in which the (transformed) parameters are \code{tparams_mean} objects. The \code{hesim_data} object provides the information required to ensure that state values are constant within groups not specified within a particular \code{stateval_tbl}; for instance, since utility only varies by health state, it is assumed constant across patients, treatment strategies, and time. The models for each cost category are combined in a list.
  
\begin{Schunk}
\begin{Sinput}
> utilitymod <- create_StateVals(utility_tbl, n = n_samples, 
+                                hesim_data = hesim_dat)
> drugcostmod <- create_StateVals(drugcost_tbl, n = n_samples,
+                                 time_reset = TRUE, hesim_data = hesim_dat)
> medcostmod <- create_StateVals(medcost_tbl, n = n_samples,
+                                hesim_data = hesim_dat)
> costmods <- list(Drug = drugcostmod,
+                  Medical = medcostmod)
\end{Sinput}
\end{Schunk}

The complete economic model is constructed by combining the transition, utility, and cost submodels.

\begin{Schunk}
\begin{Sinput}
R> ictstm <- IndivCtstm$new(trans_model = transmod,
+                           utility_model = utilitymod,
+                           cost_models = costmods)
\end{Sinput}
\end{Schunk}

Once the economic model has been constructed, it is straightforward to simulate outcomes. Trajectories through the multi-state model are simulated with the \code{$sim_disease()} method with patients assumed to live to a maximum age of 100.

\begin{Schunk}
\begin{Sinput}
R> ictstm$sim_disease(max_age = 100)
R> head(ictstm$disprog_)
\end{Sinput}
\begin{Soutput}
   sample strategy_id patient_id grp_id from to final time_start
1:      1           1          1      1    1  3     1   0.000000
2:      1           1          2      1    1  2     0   0.000000
3:      1           1          2      1    2  3     1   8.058198
4:      1           1          3      1    1  2     0   0.000000
5:      1           1          3      1    2  3     1  15.250551
6:      1           1          4      1    1  3     1   0.000000
   time_stop
1: 11.183651
2:  8.058198
3: 12.794488
4: 15.250551
5: 23.791357
6:  6.047286
\end{Soutput}
\end{Schunk}

\begin{figure}[h]
\centering
\includegraphics{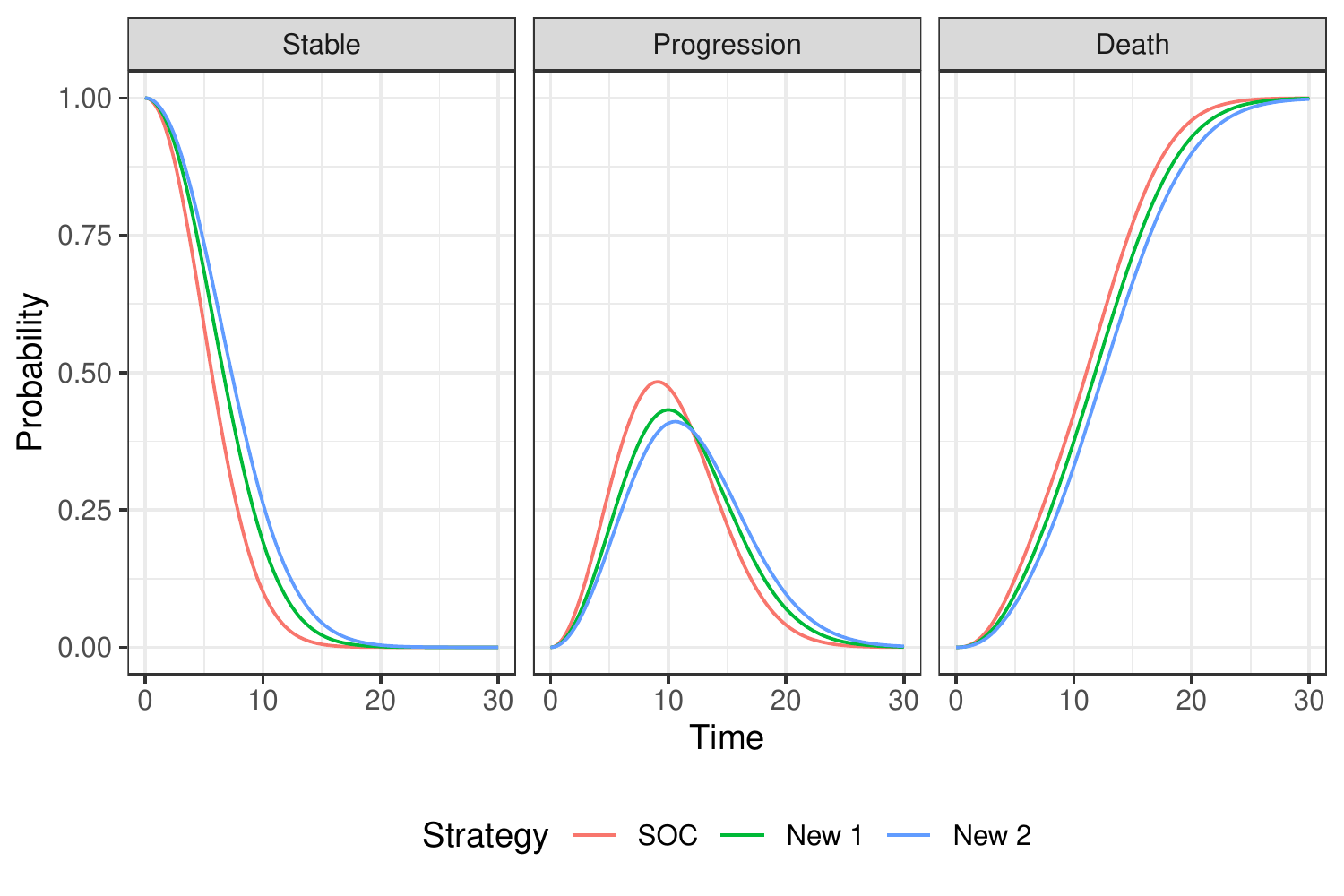}
\caption{Mean simulated state probabilities from the probabilistic sensitivity analysis.} \label{fig:ictstm-stprobs}
\end{figure}

State probabilities (averaged across patients in a given subgroup) can be constructed from the disease progression simulation output with the \code{$sim_stateprobs()} method. As shown in Figure~\ref{fig:ictstm-stprobs}, patients using the newer treatments remain in stable disease longer and have longer expected survival.  

\begin{Schunk}
\begin{Sinput}
R> ictstm$sim_stateprobs(t = seq(0, 30 , 1/12))
R> head(ictstm$stateprobs_)
\end{Sinput}
\begin{Soutput}
   sample strategy_id grp_id state_id          t  prob
1:      1           1      1        1 0.00000000 1.000
2:      1           1      1        1 0.08333333 1.000
3:      1           1      1        1 0.16666667 1.000
4:      1           1      1        1 0.25000000 0.999
5:      1           1      1        1 0.33333333 0.998
6:      1           1      1        1 0.41666667 0.996
\end{Soutput}
\end{Schunk}

\code{autoplot()} methods are available to quickly visualize simulation output using \pkg{ggplot2} graphics \citep{wickham2016ggplot2}. State probabilities from such a plot are shown in Figure~\ref{fig:ictstm-stprobs}, with labels for the treatment strategies and health states based on the output of \code{get_labels()} produced above.

\begin{Schunk}
\begin{Sinput}
R> library("ggplot2")
R> autoplot(ictstm$stateprobs_, labels = labs_indiv,
+           ci = FALSE) 
\end{Sinput}
\end{Schunk}

QALYs and costs are simulated from the disease progression simulation output using the utility and cost models with the \code{$sim_qalys()} and \code{$sim_costs()} methods. Outcomes are simulated for each discount rate specified with the \code{dr} argument. By default, \code{$sim_qalys()} will return life-years (i.e., a utility value of 1 for each year of life) in addition to QALYs. 

\begin{Schunk}
\begin{Sinput}
R> ictstm$sim_qalys(dr = c(0,.03))
R> head(ictstm$qalys_)
\end{Sinput}
\begin{Soutput}
   sample strategy_id grp_id state_id dr    qalys      lys
1:      1           1      1        1  0 4.933188 6.374648
2:      1           1      1        2  0 3.233055 5.255885
3:      1           2      1        1  0 5.604984 7.242740
4:      1           2      1        2  0 2.683823 4.363013
5:      1           3      1        1  0 6.289275 8.126979
6:      1           3      1        2  0 2.915707 4.739981
\end{Soutput}
\end{Schunk}

\begin{Schunk}
\begin{Sinput}
R> ictstm$sim_costs(dr = .03)
R> head(ictstm$costs_)
\end{Sinput}
\begin{Soutput}
   sample strategy_id grp_id state_id   dr category      costs
1:      1           1      1        1 0.03     Drug  11352.373
2:      1           1      1        2 0.03     Drug   4691.177
3:      1           2      1        1 0.03     Drug  76292.008
4:      1           2      1        2 0.03     Drug   3819.849
5:      1           3      1        1 0.03     Drug 105209.810
6:      1           3      1        2 0.03     Drug   4042.658
\end{Soutput}
\end{Schunk}

For a given subgroup, QALYs and costs are simulated for each PSA sample, treatment strategy, and health state. They are summarized with the \code{$summarize()} method, which first sums across health states and then computes a (weighted) average across patients, by PSA sample, treatment strategy, and subgroup. The output is a generic cost-effectiveness object (of class \code{ce}) that can be used to perform CEA as demonstrated in Section~\ref{sec:cea}. Here, we use the \code{summary()} method to summarize the results and make use of the ``pipes'' from the \pkg{magrittr} to nicely format the output. 

\begin{Schunk}
\begin{Sinput}
R> library("magrittr")
R> ce_sim_ictstm <- ictstm$summarize()
R> summary(ce_sim_ictstm, labels = labs_indiv) %>%
+    format()
\end{Sinput}
\begin{Soutput}
   Discount rate        Outcome                      SOC
1:          0.00          QALYs        7.87 (7.12, 8.60)
2:          0.03          QALYs        6.58 (6.03, 7.14)
3:          0.03    Costs: Drug  15,352 (14,431, 16,234)
4:          0.03 Costs: Medical  46,784 (4,518, 144,256)
5:          0.03   Costs: total 62,136 (19,888, 159,037)
                       New 1                      New 2
1:         8.59 (7.81, 9.40)         9.20 (8.42, 10.06)
2:         7.10 (6.52, 7.69)          7.52 (6.96, 8.11)
3:   78,465 (73,854, 83,200)  105,306 (99,228, 111,686)
4:   46,222 (4,543, 137,303)    46,840 (4,746, 143,093)
5: 124,687 (82,038, 218,484) 152,146 (109,625, 245,279)
\end{Soutput}
\end{Schunk}

\subsection{Partitioned survival model} \label{sec:psm-example}
In the previous section a multi-state model was fit to IPD. Partitioned survival analysis is an alternative approach that is frequently used for economic evaluations of therapies in oncology. Although there is reason to believe that PSMs are more biased than multi-state models \citep{woods2020partitioned}, PSMs are easier to parameterize with aggregate-level data because techniques for fitting multi-state models with aggregate data require more constraints and are less established \citep{price2011parameterization, pahuta2019technique, jansen2020multi}. PSMs, in contrast, consist of separate independent survival models that can be estimated with aggregate-level data by reconstructing IPD from digitized survival curves \citep{guyot2012enhanced}. 

Since a PSM is a cohort model, the \code{hesim_data} object must be modified accordingly. We now assume that the cohort can be characterized by four representative patients, each given an equal weight. We consider separate subgroups (identified with \code{grp_id}) for males and females. 

\begin{Schunk}
\begin{Sinput}
R> hesim_dat$patients <- data.table(
+    patient_id = 1:4,
+    grp_id = c(1, 1, 2, 2),
+    patient_wt = rep(1/4, 4),
+    age = c(55, 65, 55, 65),
+    female = c(1, 1, 0, 0),
+    grp_name = rep(c("Female", "Male"), 
+                   each = 2)
+  )
\end{Sinput}
\end{Schunk}

Since there are now subgroups, we generate new labels that include the subgroup variables.

\begin{Schunk}
\begin{Sinput}
R> labs_cohort <- get_labels(hesim_dat)
\end{Sinput}
\end{Schunk}

\subsubsection{Parameterization}
A multi-state dataset can be converted into a dataset with one row per patient and outcomes for PFS and OS using the function \code{as_pfs_os()}. 

\begin{Schunk}
\begin{Sinput}
R> onc_pfs_os <- as_pfs_os(onc3, patient_vars = c("patient_id", "female", "age", 
+                                                 "strategy_name"))
R> onc_pfs_os[patient_id %in% c(1, 2)]
\end{Sinput}
\begin{Soutput}
   patient_id female      age strategy_name pfs_time pfs_status
1:          1      0 59.85813         New 2 2.420226          1
2:          2      0 62.57282         New 2 7.497464          0
   os_status   os_time
1:         1 14.620258
2:         0  7.497464
\end{Soutput}
\end{Schunk}

Survival models must be fit for both PFS and OS. To imitate a realistic scenario, we assume that IPD is only available for SOC and that relative treatment effects are estimated with aggregate data (e.g., using a NMA). As such, we start by fitting Weibull models for SOC. 

One difficulty with partitioned survival analysis is that it does not account for the correlation in the survival endpoints and the curves may consequently cross during the PSA. One way to better account for correlations between the endpoints is to draw PSA samples via bootstrapping whereby the PFS and OS models are refit repeatedly to resamples of the estimation dataset. To allow for bootstrapping we create a \code{partsurvfit} object, which stores both the survival models and the estimation data.

\begin{Schunk}
\begin{Sinput}
> onc_pfs_os_soc <- onc_pfs_os[strategy_name == "SOC"]
> fit_pfs_soc_wei <- flexsurvreg(
+   Surv(pfs_time, pfs_status) ~ female,
+   data = onc_pfs_os_soc,
+   dist = "weibull")
> 
> fit_os_soc_wei <- flexsurvreg(
+   Surv(os_time, os_status) ~  female,
+   data = onc_pfs_os_soc,
+   dist = "weibull")
> 
> psmfit_soc_wei <- partsurvfit(
+   flexsurvreg_list(pfs = fit_pfs_soc_wei, os = fit_os_soc_wei),
+   data = onc_pfs_os_soc
+ )
\end{Sinput}
\end{Schunk}

The relative treatment effects are available (from a hypothetical external analysis) in terms of regression coefficients for the scale parameter on the log scale. The matrices \code{coef_pfs_wei} and \code{coef_os_wei} contain samples of the treatment effects for the PSA. We display the first 6 rows of the PFS coefficients.

\begin{Schunk}
\begin{Sinput}
R> head(coef_pfs_wei)
\end{Sinput}
\begin{Soutput}
          new1      new2
[1,] 0.1814189 0.2654431
[2,] 0.1390351 0.2907570
[3,] 0.1298460 0.2307868
[4,] 0.1729702 0.3143110
[5,] 0.1988663 0.3179721
[6,] 0.1305311 0.2261630
\end{Soutput}
\end{Schunk}

The parameters for utility and medical costs remain the same as in the previous section. We do, however, modify drug costs. Since we are no longer using an individual-level model, costs cannot vary as a function of time in a health state. We now assume conservatively that drug costs remain at their initial level when in the progression state.  

\begin{Schunk}
\begin{Sinput}
R> drugcost_tbl <- stateval_tbl(
+    drugcost_dt[time_start ==  0][, time_start := NULL],
+    dist = "fixed")
R> print(drugcost_tbl)
\end{Sinput}
\begin{Soutput}
   strategy_id state_id   est
1:           1        1  2000
2:           1        2  1500
3:           2        1 12000
4:           2        2  1500
5:           3        1 15000
6:           3        2  1500
\end{Soutput}
\end{Schunk}

\subsubsection{Simulation}
As with  the multi-state model, input data for the survival models is constructed with the \code{expand()} function.

\begin{Schunk}
\begin{Sinput}
R> survmods_data <- expand(hesim_dat, by = c("strategies", "patients"))
R> head(survmods_data)
\end{Sinput}
\begin{Soutput}
   strategy_id patient_id strategy_name soc new1 new2 grp_id
1:           1          1           SOC   1    0    0      1
2:           1          2           SOC   1    0    0      1
3:           1          3           SOC   1    0    0      2
4:           1          4           SOC   1    0    0      2
5:           2          1         New 1   0    1    0      1
6:           2          2         New 1   0    1    0      1
   patient_wt age female grp_name
1:       0.25  55      1   Female
2:       0.25  65      1   Female
3:       0.25  55      0     Male
4:       0.25  65      0     Male
5:       0.25  55      1   Female
6:       0.25  65      1   Female
\end{Soutput}
\end{Schunk}

Since we did not fit a survival regression model containing both the baseline risk and relative treatment effects, the survival parameter object must be constructed manually. We obtain the parameters of the fitted Weibull regression models with the generic function \code{create_params()}, and then manually add the relative treatment effects to the coefficients for the scale parameters. (Note that if we had not fit a model for SOC, the survival parameter objects could have still been constructed with \code{params_surv()}.) 

The parameters for SOC are drawn by bootstrapping. Although this better accounts for the correlation between PFS and OS, it is slower than drawing the parameters from a multivariate normal distribution (as employed in Section~\ref{sec:ictstm-example}) because the survival models are refit for each PSA sample. We use the \code{max_errors} argument to allow for errors during some of the bootstrap replications (e.g., if a model fails to converge).

\begin{Schunk}
\begin{Sinput}
R> survmods_params <- create_params(psmfit_soc_wei, n = n_samples, 
+                                   uncertainty = "bootstrap", 
+                                   max_errors = 5, silent = TRUE)
R> survmods_params$pfs$coefs$scale <- cbind(survmods_params$pfs$coefs$scale,
+                                           coef_pfs_wei)
R> survmods_params$os$coefs$scale <- cbind(survmods_params$os$coefs$scale,
+                                          coef_os_wei)
R> survmods_params$pfs$coefs$scale[1:2, ]
\end{Sinput}
\begin{Soutput}
     (Intercept)     female      new1      new2
[1,]    1.784666 -0.1364781 0.1814189 0.2654431
[2,]    1.776743 -0.1173777 0.1390351 0.2907570
\end{Soutput}
\end{Schunk}

Survival curves for both PFS and OS are predicted using the \code{PsmCurves} class, which is setup using the input data and the parameters of the Weibull model. However, since we manually created the parameter object---meaning that the model terms were not generated using a formula interface---the input data must contain each term contained in the parameter object. We must consequently create a column of ones for the intercept.

\begin{Schunk}
\begin{Sinput}
R> survmods_data[, ("(Intercept)") := 1]
R> survmods <- create_PsmCurves(survmods_params, 
+                               input_data = survmods_data)
\end{Sinput}
\end{Schunk}

We re-instantiate all of the utility and cost models. This is necessary because our patient population changed and we are now only making predictions for 4 representative patients (rather than 1,000 in the individual simulation). The \code{hesim_data} argument to \code{create_StateVals()} does this for us automatically. 

\begin{Schunk}
\begin{Sinput}
R> utilitymod <- create_StateVals(utility_tbl, n = n_samples, 
+                                 hesim_data = hesim_dat)
R> drugcostmod <- create_StateVals(drugcost_tbl, n = n_samples, 
+                                  hesim_data = hesim_dat)
R> medcostmod <- create_StateVals(medcost_tbl, n = n_samples, 
+                                 hesim_data = hesim_dat)
R> costmods <- list(Drug = drugcostmod, Medical = medcostmod)
\end{Sinput}
\end{Schunk}

Now that the survival, utility, and cost submodels have been constructed, we can instantiate the complete PSM.  

\begin{Schunk}
\begin{Sinput}
R> psm <- Psm$new(survival_models = survmods,
+                 utility_model = utilitymod,
+                 cost_models = costmods)
\end{Sinput}
\end{Schunk}

\begin{figure}[h]
\centering
\includegraphics{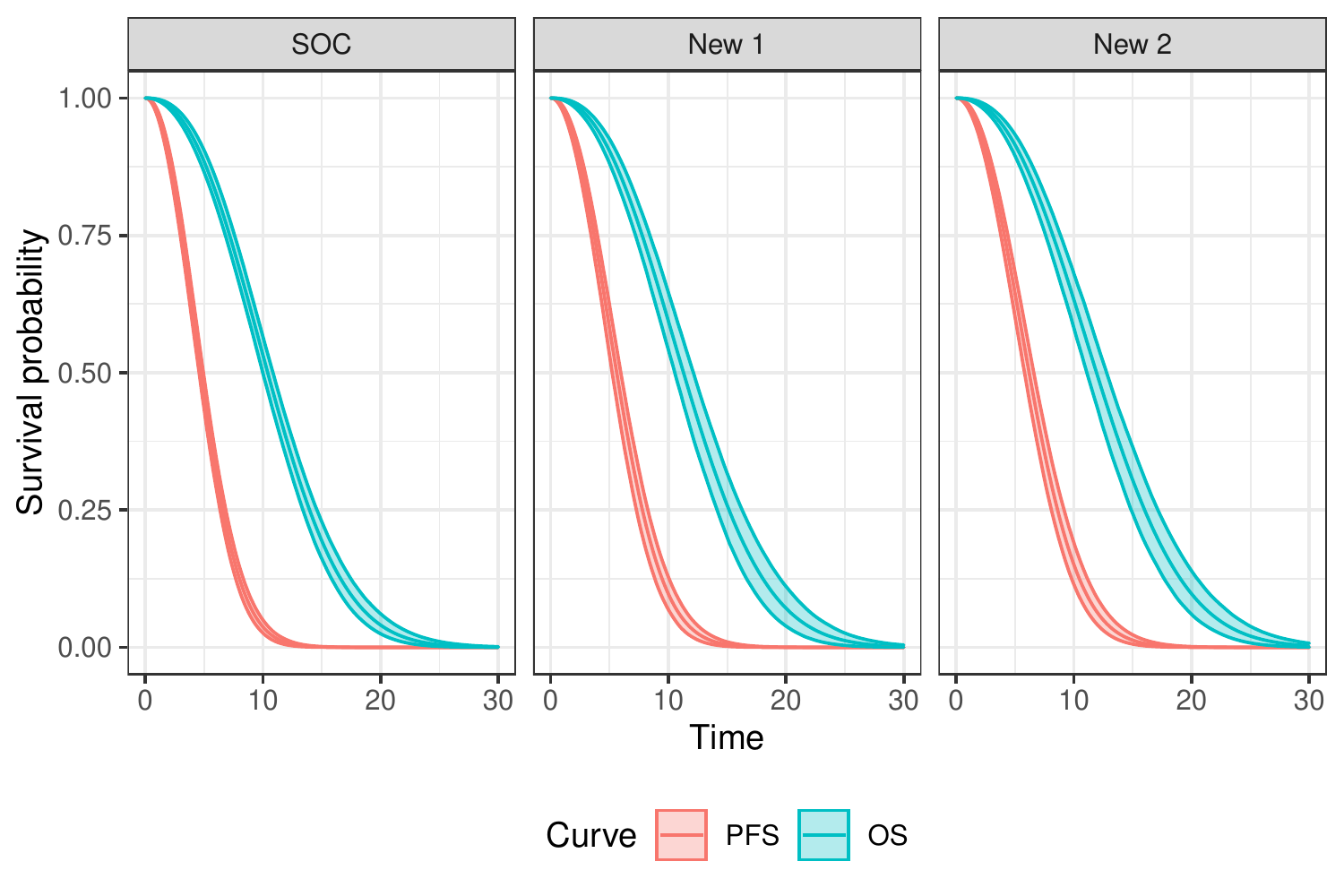}
\caption{Mean progression free survival and overall survival, weighted by the weight given to each patient in the target population. The solid line and shaded regions represent means and 95\% confidence intervals from the probabilistic sensitivity analysis, respectively.} \label{fig:psm-survival}
\end{figure}

We first simulate survival curves by month for 30 years. Survival curves produced using \code{autoplot()} are displayed in Figure~\ref{fig:psm-survival}. 

\begin{Schunk}
\begin{Sinput}
R> times <- seq(0, 30, by = .1)
R> psm$sim_survival(t = times)
R> head(psm$survival_)
\end{Sinput}
\begin{Soutput}
   sample strategy_id patient_id grp_id patient_wt curve   t
1:      1           1          1      1       0.25     1 0.0
2:      1           1          1      1       0.25     1 0.1
3:      1           1          1      1       0.25     1 0.2
4:      1           1          1      1       0.25     1 0.3
5:      1           1          1      1       0.25     1 0.4
6:      1           1          1      1       0.25     1 0.5
    survival
1: 1.0000000
2: 0.9998540
3: 0.9993126
4: 0.9982992
5: 0.9967670
6: 0.9946815
\end{Soutput}
\end{Schunk}

\begin{Schunk}
\begin{Sinput}
R> autoplot(psm$survival_, 
+           labels = c(labs_cohort,
+                      list(curve = c("PFS" = 1, "OS" = 2))
+                      ),
+           ci = TRUE, ci_style = "ribbon")
\end{Sinput}
\end{Schunk}

State probabilities are easily computed from these survival curves with the \code{$sim_stateprobs()} method. The area under the PFS curve is the proportion of patients in stable disease while the area between the PFS and OS curves is the proportion of patients whose disease has progressed.

\begin{Schunk}
\begin{Sinput}
R> psm$sim_stateprobs()
R> psm$stateprobs_[sample == 1 & patient_id == 1 & state_id == 2 & t == 12]
\end{Sinput}
\begin{Soutput}
   sample strategy_id patient_id grp_id patient_wt state_id  t
1:      1           1          1      1       0.25        2 12
2:      1           2          1      1       0.25        2 12
3:      1           3          1      1       0.25        2 12
        prob
1: 0.3215069
2: 0.3934037
3: 0.4148810
\end{Soutput}
\end{Schunk}

QALYs and costs are simulated using the same \code{sim_qalys()} and \code{sim_costs()} methods as in the previous section, and the results are summarized in the same fashion as well. A difference, however, is that we now use the \code{by_grp = TRUE} option to summarize costs and QALYs by subgroup (rather than aggregating across all groups). 

\begin{Schunk}
\begin{Sinput}
R> psm$sim_qalys(dr = .03) 
R> psm$sim_costs(dr = .03)
\end{Sinput}
\end{Schunk}

\begin{Schunk}
\begin{Sinput}
R> ce_sim_psm <- psm$summarize(by_grp = TRUE)
R> summary(ce_sim_psm, labels = labs_cohort) %>%
+    format()
\end{Sinput}
\begin{Soutput}
    Group Discount rate        Outcome                      SOC
1: Female          0.03          QALYs        6.10 (5.55, 6.63)
2: Female          0.03    Costs: Drug  15,225 (14,629, 15,900)
3: Female          0.03 Costs: Medical  52,296 (4,435, 180,774)
4: Female          0.03   Costs: total 67,521 (19,657, 195,939)
5:   Male          0.03          QALYs        6.51 (5.99, 7.08)
6:   Male          0.03    Costs: Drug  16,247 (15,553, 17,004)
7:   Male          0.03 Costs: Medical  52,677 (4,767, 171,325)
8:   Male          0.03   Costs: total 68,924 (20,902, 187,307)
                       New 1                      New 2
1:         6.61 (6.00, 7.22)          6.96 (6.36, 7.60)
2:   66,473 (62,592, 70,188)    88,466 (83,561, 93,548)
3:   52,717 (4,871, 171,715)    53,138 (5,156, 169,727)
4: 119,189 (70,613, 239,546)  141,604 (92,445, 260,157)
5:         7.05 (6.46, 7.70)          7.42 (6.81, 8.07)
6:   74,027 (69,772, 78,081)   98,691 (93,156, 104,387)
7:   52,709 (5,188, 165,850)    52,874 (5,417, 162,113)
8: 126,736 (78,810, 238,080) 151,565 (103,556, 262,961)
\end{Soutput}
\end{Schunk}

\subsection{Cohort disctete time state transition model} \label{sec:cdtstm-example}
The third model type supported by \pkg{hesim} is the cDTSTM. Although it is commonly used in health economics, the approaches taken in Section~\ref{sec:ictstm-example} and \ref{sec:psm-example} make better use of the data if the time-to-event outcomes are continuously observed. Unfortunately, observations are sometimes only collected at arbitrary times, which results in what is referred to as panel data. This may, for instance, occur in routine practice oncology data when progression of disease is not continuously evaluated. To facilitate analysis of this type of data, we have converted the \code{onc3} dataset from above into panel form. 

\begin{Schunk}
\begin{Sinput}
R> head(onc3p)
\end{Sinput}
\begin{Soutput}
         state strategy_name female      age patient_id      time
1:      Stable         New 2      0 59.85813          1  0.000000
2: Progression         New 2      0 59.85813          1  2.420226
3:       Death         New 2      0 59.85813          1 14.620258
4:      Stable         New 2      0 62.57282          2  0.000000
5:      Stable         New 2      0 62.57282          2  7.497464
6:      Stable           SOC      1 61.44379          3  0.000000
   strategy_id state_id
1:           3        1
2:           3        2
3:           3        3
4:           3        1
5:           3        1
6:           1        1
\end{Soutput}
\end{Schunk}

\subsubsection{Parameterization}
As in Section~\ref{sec:psm-example}, we consider a case where IPD is available for SOC but aggregate-level data is used to estimate the relative treatment effects. When exact transition times are not observed, the survival models estimated in Section~\ref{sec:ictstm-example} are inappropriate. The \pkg{msm} package can, however, be used to estimate models from such data. The loss of information does come at a price as only exponential and piecewise exponential models are possible. We illustrate by fitting an exponential multi-state model for SOC using \pkg{msm}. (Note that if exact times were observed and we set the argument \code{exacttimes = TRUE}, then the parameter estimates would be the same as an exponential model fit using \code{flexsurvreg()}.)

\begin{Schunk}
\begin{Sinput}
R> library("msm")
R> qinit <- matrix(0, nrow = 3, ncol = 3)
R> qinit[1, 2] <-  0.28; qinit[1, 3] <-  0.013; qinit[2, 3] <-  0.10
R> msm_fit <- msm(state_id ~ time, subject = patient_id, 
+                 data = onc3p[strategy_name == "SOC"],
+                 exacttimes = FALSE,
+                 covariates = ~ age + female,
+                 qmatrix = qinit, gen.inits = FALSE)
\end{Sinput}
\end{Schunk}

We assume that the relative treatment effects were estimated as relative risks. Either a posterior distribution from a Bayesian analysis or parameters of a parametric distribution can be used. We use the latter approach here and assume that the relative risk is approximately normally distributed on the log scale. \pkg{hesim} provides the \code{define_rng()} and \code{eval_rng()} functions for defining and evaluating random number generators for a PSA. While standard functions in \proglang{R} could be used to draw random deviates from probability distributions instead, these functions can be more convenient for health economic modeling. Notable advantages include only needing to specify the number of random samples once, being able to pass moments of distributions as arguments (as in \code{stateval_tbl()}), and guaranteeing a common output (either a vector or a two-dimensional tabular object, with dimensions determined by the number or random samples). 

\begin{Schunk}
\begin{Sinput}
R> params_rr <- list(
+    lrr_12_est = c(soc = log(1), new1 = log(.80), new2 = log(.71)),
+    lrr_12_se = c(soc = 0, new1 = .03, new2 = .04),
+    lrr_13_est = c(soc = log(1), new1 = log(.90), new2 = log(.85)),
+    lrr_13_se = c(soc = 0, new1 = .02, new2 = .03)
+  )  
R> 
R> params_rr_def <- define_rng({
+    list(
+      rr_12 = lognormal_rng(lrr_12_est, lrr_12_se),
+      rr_13 = lognormal_rng(lrr_13_est, lrr_13_se)
+    )}, n = n_samples)
R> params_rr_rng <- eval_rng(params_rr_def, params_rr)
\end{Sinput}
\end{Schunk}

In Sections~\ref{sec:ictstm-example} and \ref{sec:psm-example}, we constructed the disease model using parameter (i.e., \code{params_surv}) objects. Here, we will use a transformed parameter (i.e., \code{tparams_transprobs}) object, which is a very flexible way to construct the transition probability matrices for a cDTSTM. A \code{tparams_transprobs} object requires a 3-dimensional array of transition matrices, where the transition matrices are ordered by PSA sample, treatment strategy, patient, and optionally (in a time-inhomogeneous model) time interval.

We start by predicting transition intensity matrices for SOC from the exponential multi-state model using the same input data used for the survival models in the PSM example. A prediction is made for each representative patient and samples of the transition intensity matrices are drawn for the PSA using an asymptotic normal approximation. The array contains $B \times N_p$ matrices where $B$ is the number of PSA samples and $N_p$ is the number of patients. 

\begin{Schunk}
\begin{Sinput}
R> transmod_data <- survmods_data
R> qmat_soc <- qmatrix(msm_fit, 
+                      newdata = transmod_data[strategy_name == "SOC"],
+                      uncertainty = "normal", n = n_samples)
R> dim(qmat_soc)
\end{Sinput}
\begin{Soutput}
[1]    3    3 4000
\end{Soutput}
\begin{Sinput}
R> qmat_soc[,, 1]
\end{Sinput}
\begin{Soutput}
           [,1]       [,2]       [,3]
[1,] -0.3533989  0.3371267 0.01627226
[2,]  0.0000000 -0.1190753 0.11907530
[3,]  0.0000000  0.0000000 0.00000000
\end{Soutput}
\end{Schunk}

Transition probability matrices of arbitrary duration are derived from the transition intensity matrices with the matrix exponential. We use a cycle length of 3 months.

\begin{Schunk}
\begin{Sinput}
R> cycle_len <- 1/4
R> pmat_soc <- expmat(qmat_soc, t = cycle_len)
R> pmat_soc[,, 1]
\end{Sinput}
\begin{Soutput}
          [,1]       [,2]       [,3]
[1,] 0.9154407 0.07945955 0.00509979
[2,] 0.0000000 0.97066990 0.02933010
[3,] 0.0000000 0.00000000 1.00000000
\end{Soutput}
\end{Schunk}

Identifiers for the complete array of transitions matrices (inclusive of SOC and the new interventions) are generated with \code{tpmatrix_id()}.

\begin{Schunk}
\begin{Sinput}
R> tpmat_id <- tpmatrix_id(transmod_data, n_samples)
R> head(tpmat_id)
\end{Sinput}
\begin{Soutput}
   sample strategy_id patient_id grp_id patient_wt
1:      1           1          1      1       0.25
2:      1           1          2      1       0.25
3:      1           1          3      2       0.25
4:      1           1          4      2       0.25
5:      1           2          1      1       0.25
6:      1           2          2      1       0.25
\end{Soutput}
\end{Schunk}

Relative risks can be predicted for each row of the input data and each sample of the coefficients for the PSA using matrix multiplication. In particular, let $x$ be a $N \times k$ matrix where $N$ is the number of rows of the input data and $k$ is the number of predictors, and $\beta$ be a $B \times k$ matrix where each row contains a given sample for the PSA. The function \code{xbeta} performs the matrix multiplication operation $x\beta^T$ to produce a $N \times B$ matrix of predicted relative risks and then ``flattens'' the matrix to create a single vector of length $N \times B$.

\begin{Schunk}
\begin{Sinput}
R> xbeta <- function(x, beta) c(x %*% t(beta))
\end{Sinput}
\end{Schunk}

We use this function to predict distributions of relative risks (for the transitions stable $\rightarrow$ progression and stable $\rightarrow$ death) for each row of the input data as a function of the three treatment strategies. 

\begin{Schunk}
\begin{Sinput}
> x_rr <- as.matrix(transmod_data[, .(soc, new1, new2)])
> rr <- cbind(xbeta(x_rr, params_rr_rng$rr_12),
+             xbeta(x_rr, params_rr_rng$rr_13))
> head(rr) 
\end{Sinput}
\begin{Soutput}
         [,1]      [,2]
[1,] 1.000000 1.0000000
[2,] 1.000000 1.0000000
[3,] 1.000000 1.0000000
[4,] 1.000000 1.0000000
[5,] 0.833283 0.9039595
[6,] 0.833283 0.9039595
\end{Soutput}
\end{Schunk}

Relative risks are applied to the elements (stable $\rightarrow$ progression (1, 2) and stable $\rightarrow$ death (1, 3)) of each transition probability matrix for SOC with \code{apply_rr()}. The result is an array of matrices for each PSA sample, treatment strategy,  and representative patient, as desired. The \code{tparams_transprobs} object can then be created by simply combining the transition probability matrices and the identification dataset. 

\begin{Schunk}
\begin{Sinput}
R> pmat <- apply_rr(pmat_soc, rr = rr,
+                   index = list(c(1, 2), c(1, 3)))
R> tprobs <- tparams_transprobs(pmat, tpmat_id)
\end{Sinput}
\end{Schunk}

\subsubsection{Simulation}
Once the model has been parameterized, the simulation is straightforward. The transition model depends only on the transition probability matrices and the cycle length,

\begin{Schunk}
\begin{Sinput}
R> transmod <- CohortDtstmTrans$new(params = tprobs, 
+                                   cycle_length = cycle_len)
\end{Sinput}
\end{Schunk}

and the complete economic model is again instantiated from the transition, utility, and cost submodels. 

\begin{Schunk}
\begin{Sinput}
R> cdtstm <- CohortDtstm$new(trans_model = transmod,
+                            utility_model = psm$utility_model,
+                            cost_models = psm$cost_models)
\end{Sinput}
\end{Schunk}

We simulate state probabilities for 120 cycles (i.e., 30 years).

\begin{Schunk}
\begin{Sinput}
R> cdtstm$sim_stateprobs(n_cycles = 30/cycle_len)
\end{Sinput}
\end{Schunk}

QALYs and costs are again simulated using a 3 percent discount rate

\begin{Schunk}
\begin{Sinput}
R> cdtstm$sim_qalys(dr = .03)
R> cdtstm$sim_costs(dr = .03)
\end{Sinput}
\end{Schunk}

and the results are summarized. The \code{by_grp} argument of \code{$summarize()} is by default \code{FALSE} so costs and QALYs are aggregated across subgroups here. 

\begin{Schunk}
\begin{Sinput}
R> ce_sim_cdtstm <- cdtstm$summarize()
R> summary(ce_sim_cdtstm, labels = labs_cohort) %>%
+    format()
\end{Sinput}
\begin{Soutput}
   Discount rate        Outcome                      SOC
1:          0.03          QALYs        5.56 (4.78, 6.26)
2:          0.03    Costs: Drug  13,862 (12,563, 14,711)
3:          0.03 Costs: Medical  62,515 (4,186, 217,838)
4:          0.03   Costs: total 76,378 (18,215, 231,236)
                       New 1                     New 2
1:         5.92 (5.10, 6.67)         6.14 (5.27, 6.84)
2:   46,196 (41,201, 50,104)   60,714 (53,418, 66,584)
3:   62,011 (4,351, 223,642)   61,636 (4,671, 213,904)
4: 108,207 (50,113, 267,630) 122,350 (64,868, 274,308)
\end{Soutput}
\end{Schunk}

\section{Cost-effectiveness analysis} \label{sec:cea}
CEA is based on the net monetary benefit (NMB). For a given parameter set $\theta$, the NMB with treatment $j$ is computed as the difference between the monetized health gains from an intervention less costs, or,

\begin{equation}
NMB(j,\theta) = e_{j}(\theta)\cdot k- c_{j}(\theta),
\end{equation}

where $e_{j}$ and $c_{j}$ are measures of health outcomes (e.g., QALYs) and costs using treatment $j$ respectively, and $k$ is a decision makers willingness to pay (WTP) per unit of a health outcome. The optimal treatment is the one that maximizes the expected NMB,

\begin{equation}
j^{*} = \argmax_j E_{\theta} \left[NMB(j,\theta)\right].
\end{equation}

For a pairwise comparison, treatment $1$ is preferred to treatment $0$ if the expected incremental net monetary benefit (INMB) is positive; that is, if $E_\theta \left[INMB\right] > 0$ where the INMB is given by

\begin{equation}
INMB(\theta) = NMB(j = 1, \theta) - NMB(j = 0, \theta).
\end{equation}

Treatments can be compared in an equivalent manner using the incremental cost-effectiveness ratio (ICER). The most common case occurs when a new treatment is more effective and more costly so that treatment $1$ is preferred to treatment $0$ if the ICER is less than than the willingness to pay threshold $k$,

\begin{equation}
k > \frac{E_\theta[c_{1} - c_{0}]}{E_\theta[e_{1} - e_{0}]} = ICER.
\end{equation}

There are three additional cases. Treatment $1$ is considered to \emph{dominate} treatment $0$ if it is more effective and less costly. Treatment $1$ is \emph{dominated} by treatment $0$ if it is less effective and more costly. Finally, treatment $1$ is preferred to treatment $0$ if it is less effective and less costly when $k < ICER$. 

\subsection{Application}
Uncertainty over $\theta$ is captured using the PSA. The simulated distribution of QALYs and costs for each treatment strategy can be produced with the \code{$summarize()} method as was demonstrated in Section~\ref{sec:illustrative-example}. We will use the output from the iCTSTM in this example. 

\begin{Schunk}
\begin{Sinput}
R> ce_sim_ictstm
\end{Sinput}
\begin{Soutput}
$costs
      category   dr sample strategy_id     costs grp_id
   1:     Drug 0.03      1           1  16043.55      1
   2:     Drug 0.03      1           2  80111.86      1
   3:     Drug 0.03      1           3 109252.47      1
   4:     Drug 0.03      2           1  14645.16      1
   5:     Drug 0.03      2           2  77962.49      1
  ---                                                  
8996:    total 0.03    999           2  95024.16      1
8997:    total 0.03    999           3 122357.69      1
8998:    total 0.03   1000           1  52638.13      1
8999:    total 0.03   1000           2 119410.47      1
9000:    total 0.03   1000           3 151164.45      1

$qalys
        dr sample strategy_id    qalys grp_id
   1: 0.00      1           1 8.166243      1
   2: 0.00      1           2 8.288806      1
   3: 0.00      1           3 9.204982      1
   4: 0.00      2           1 7.440970      1
   5: 0.00      2           2 8.414148      1
  ---                                        
5996: 0.03    999           2 7.134710      1
5997: 0.03    999           3 7.399217      1
5998: 0.03   1000           1 6.435002      1
5999: 0.03   1000           2 7.103850      1
6000: 0.03   1000           3 7.584523      1

attr(,"class")
[1] "ce"
\end{Soutput}
\end{Schunk}

A CEA is performed from this output with the \code{cea()} and \code{cea_pw()} functions. \code{cea()} summarizes results by simultaneously accounting for each treatment strategy in the analysis, while \code{cea_pw()} summarizes ``pairwise'' results in which each treatment is compared to a comparator (in this case SOC). 

\begin{Schunk}
\begin{Sinput}
> wtp <- seq(0, 250000, 500) 
> cea_ictstm <- cea(ce_sim_ictstm, dr_costs = .03, dr_qalys = .03, k = wtp)
> cea_pw_ictstm <- cea_pw(ce_sim_ictstm, comparator = 1,
+                         dr_qalys = .03, dr_costs = .03,
+                         k = wtp)
\end{Sinput}
\end{Schunk}

\subsubsection{Incremental cost-effectiveness ratio}
The \code{cea_pw()} function computes the ICER. A summary table is produced with \code{icer()} and \code{format()} formats the table for pretty printing. The argument \code{k} is the WTP threshold and is used to compute the INMB. Estimates of incremental QALYs and incremental costs are computed by averaging over PSA samples. By default, 95\% confidence intervals are presented and are computed using quantiles from the PSA.

\begin{Schunk}
\begin{Sinput}
R> icer(cea_pw_ictstm, k = 100000, labels = labs_indiv) %>%
+    format()
\end{Sinput}
\begin{Soutput}
             Outcome                     New 1
1: Incremental QALYs         0.52 (0.16, 0.90)
2: Incremental costs   62,551 (51,685, 70,389)
3:   Incremental NMB -11,003 (-42,132, 21,247)
4:              ICER                   121,345
                      New 2
1:        0.93 (0.58, 1.27)
2: 90,010 (77,342, 100,581)
3:  3,223 (-28,582, 33,456)
4:                   96,543
\end{Soutput}
\end{Schunk}

\subsubsection{Representing decision uncertainty}
\begin{figure}[t!]
\centering
\begin{subfigure}[b]{.49\textwidth}
\centering
\includegraphics{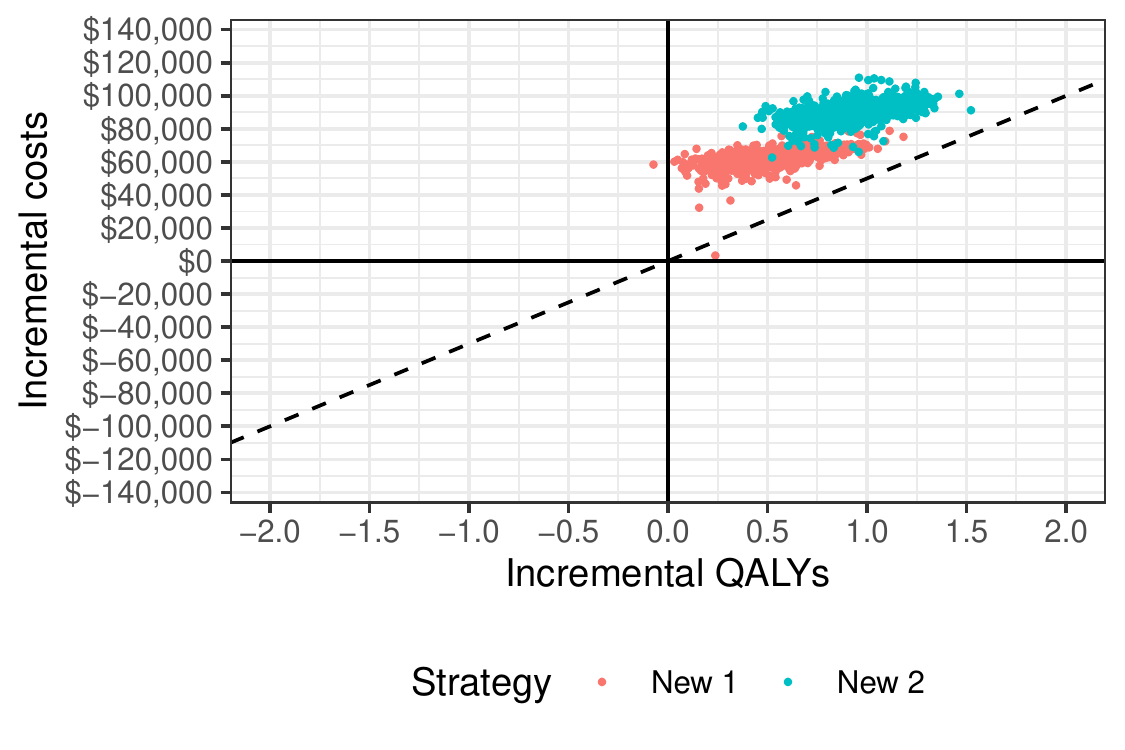}
\caption{Cost-effectiveness plane} \label{subfig:ce-plane}
\end{subfigure}
\begin{subfigure}[b]{.49\textwidth}
\centering
\includegraphics{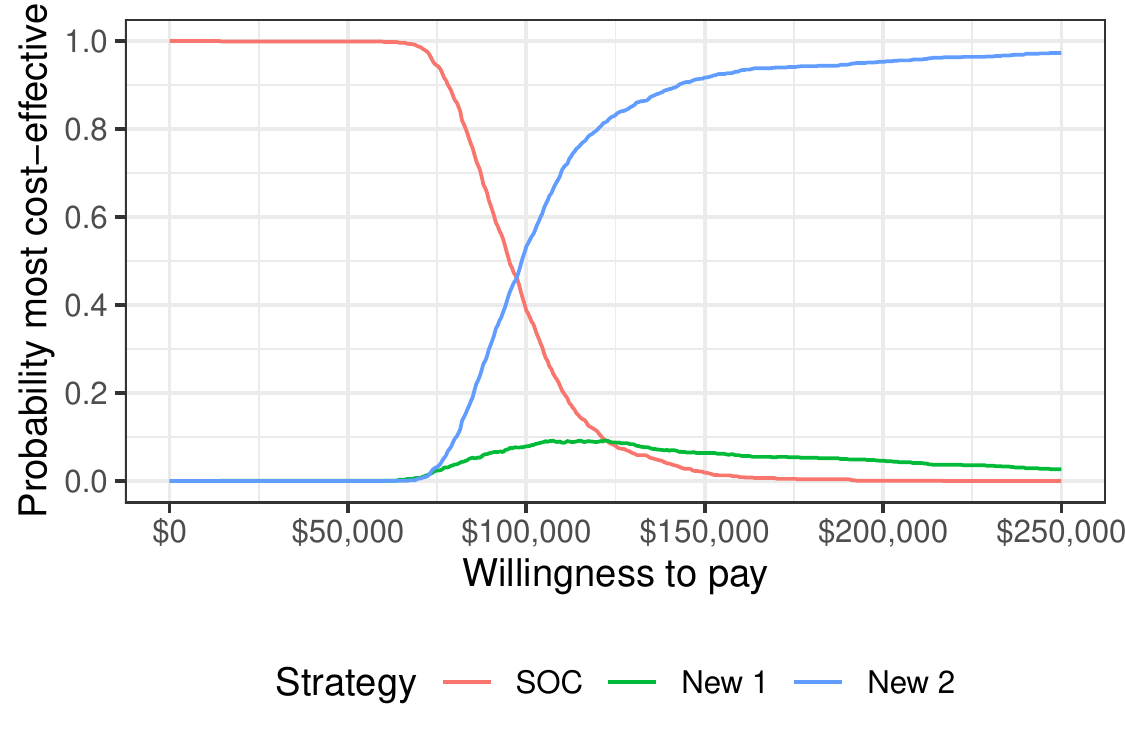}
\caption{CEAC (simultaneous)} \label{subfig:ceac-simultaneous}
\end{subfigure}
\par\bigskip 
\begin{subfigure}[b]{.49\textwidth}
\centering
\includegraphics{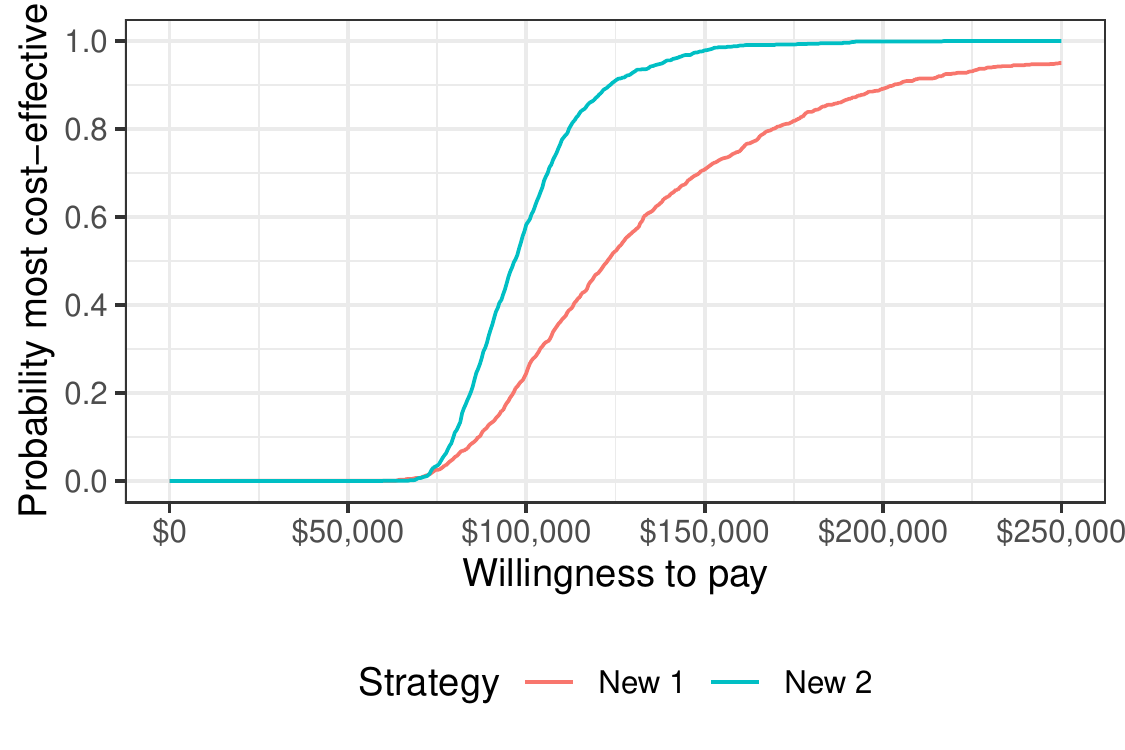}
\caption{CEAC (pairwise)} \label{subfig:ceac-pw}
\end{subfigure}
\begin{subfigure}[b]{.49\textwidth}
\centering
\includegraphics{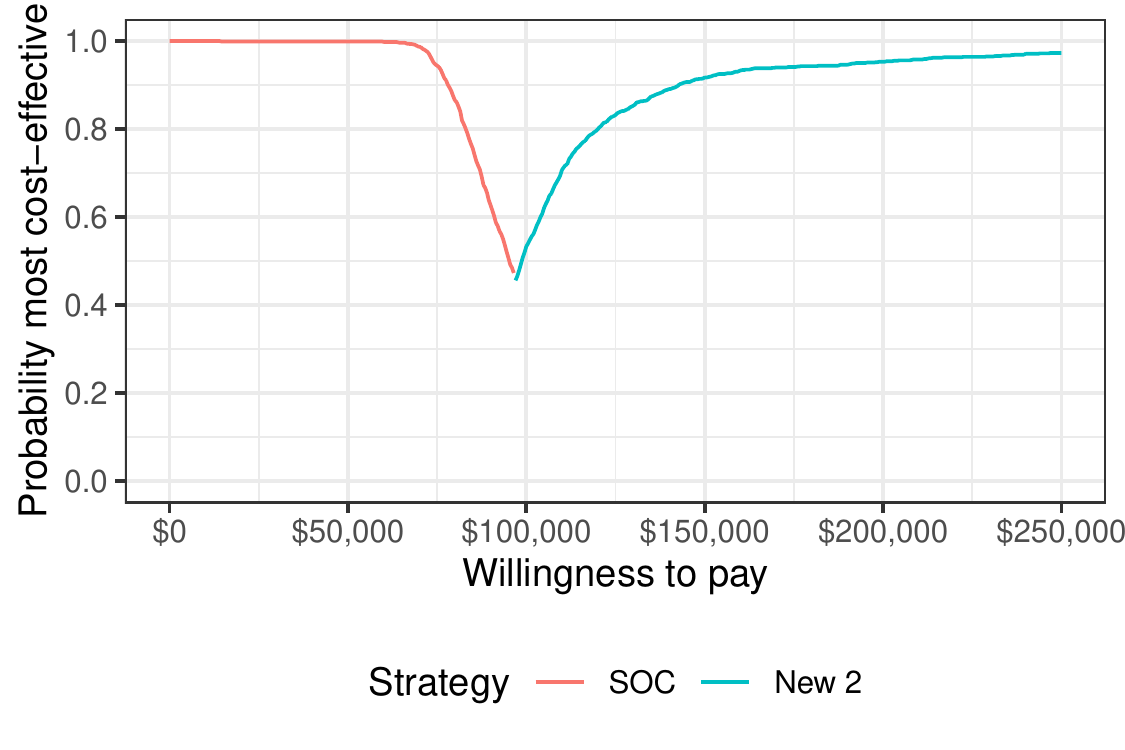}
\caption{CEAF} \label{subfig:ceaf}
\end{subfigure}
\par\bigskip 
\begin{subfigure}[b]{.49\textwidth}
\centering
\includegraphics{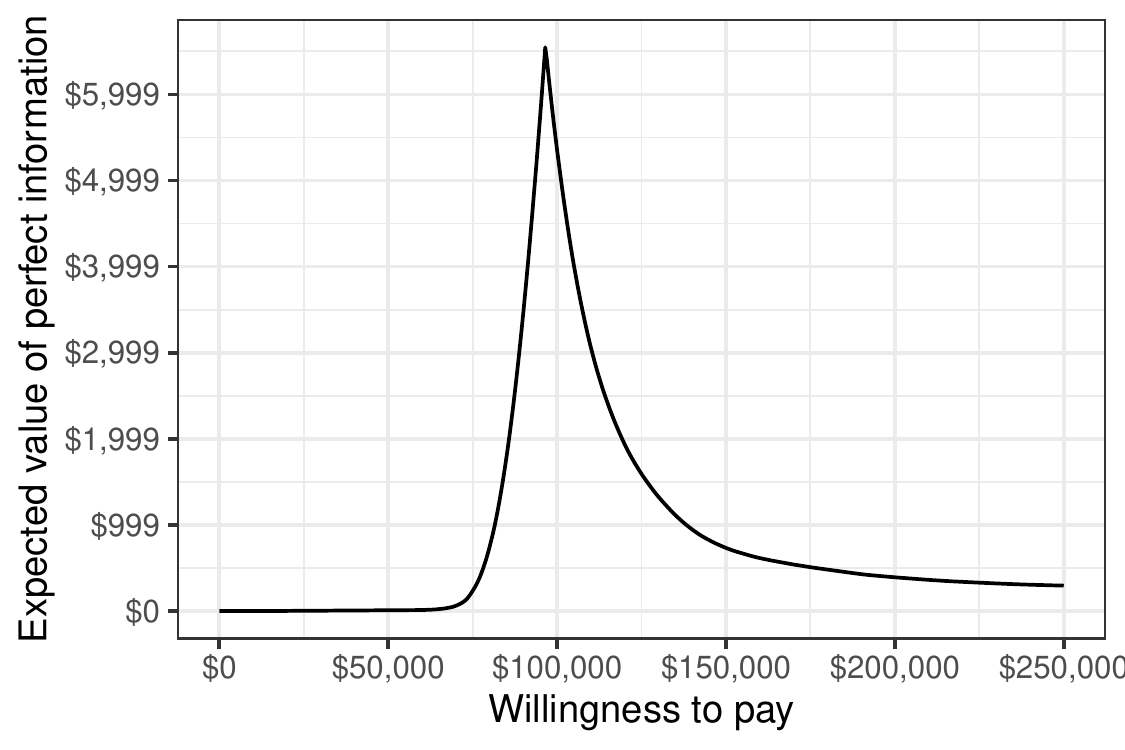}
\caption{EVPI} \label{subfig:evpi}
\end{subfigure}
\caption{Representations of decision uncertainty.}\label{fig:cea-plots}
\end{figure}

A number of common measures can be used to represent decision uncertainty from a CEA, including cost-effectiveness planes, cost-effectiveness acceptability curves (CEACs), cost-effectiveness acceptability frontiers (CEAFs), and the expected value of perfect information (EVPI). Plots using \pkg{ggplot2} can be quickly generated with the \code{plot_ceplane()}, \code{plot_ceac()}, \code{plot_ceaf()}, and \code{plot_evpi()} functions, respectively. Output from \code{cea()} and \code{cea_pw()} is readily available, so users may create custom plots as well. Example figures from the code that follows are shown in Figure~\ref{fig:cea-plots}.

The cost-effectiveness plane plots the incremental effectiveness of a treatment strategy (relative to a comparator) against the incremental cost of the treatment strategy. The plot is useful because it demonstrates both the uncertainty and the magnitude of the estimates. Each point on the plot is from a particular draw from the PSA. Data for plotting a cost-effectiveness plane comes from the \code{delta} output generated from \code{cea_pw()}.

\begin{Schunk}
\begin{Sinput}
R> plot_ceplane(cea_pw_ictstm, labels = labs_indiv)
\end{Sinput}
\end{Schunk}

The CEAC can be generated based on simultaneous comparisons of all treatment strategies or by comparing each treatment strategy to a single comparator. The former is the probability that each strategy is the most cost-effective and is available from the \code{mce} element produced by \code{cea()}. The relevant information for the latter is available from the \code{ceac} element produced by \code{cea_pw()}, and is the probability that each treatment is more cost-effective than the comparator.

\begin{Schunk}
\begin{Sinput}
R> plot_ceac(cea_ictstm, labels = labs_indiv)
\end{Sinput}
\begin{Sinput}
R> plot_ceac(cea_pw_ictstm, labels = labs_indiv)
\end{Sinput}
\end{Schunk}

One drawback of the CEAC is that the probability of being cost-effective cannot be used to determine the optimal treatment option. Instead, if the objective is to maximize health gain, then decisions should be based on the expected NMB. The CEAF, which plots the probability that the optimal treatment strategy (i.e., the strategy with the highest expected NMB) is cost-effective, is appropriate in this context. A CEAF curve can be created by using the \code{best} column from the \code{mce} element to subset to the treatment strategy with the highest expected NMB for each WTP value.

\begin{Schunk}
\begin{Sinput}
R> plot_ceaf(cea_ictstm, labels = labs_indiv)
\end{Sinput}
\end{Schunk}

A limitation of CEACs and CEAFs is that they ignore the magnitude of QALY or cost gains. A measure which combines the probability of being most effective with the magnitude of the expected NMB is the EVPI. Intuitively, the EVPI provides an estimate of the amount that a decision maker would be willing to pay to collect additional data and completely eliminate uncertainty. Mathematically, the EVPI is defined as the difference between the maximum expected NMB given perfect information and the maximum expected NMB given current information. In other words, we calculate the NMB for the optimal treatment strategy for each random draw of the parameters and compare that to the NMB for the treatment strategy that is optimal when averaging across all parameters,

\begin{equation}
EVPI = E_\theta \left[\max_j NMB(j, \theta)\right] - \max_j E_\theta \left [ NMB(j, \theta)\right].
\end{equation}

The \code{cea()} function performs the EVPI calculation across all simulation draws from the PSA and for a number of WTP values. The kink in the plot represents the WTP value where the optimal strategy changes.

\begin{Schunk}
\begin{Sinput}
R> plot_evpi(cea_ictstm, labels = labs_indiv)
\end{Sinput}
\end{Schunk}

\section{Discussion} \label{sec:discussion}
This paper formally describes the economic models and cost-effectiveness framework used by the \pkg{hesim} package. An illustrative example is provided for a three state model commonly used to evaluate treatments in oncology. However, since \pkg{hesim} aims to be flexible, the example only covers a subset of the relevant modeling approaches and functionality. Additional examples as well as documentation for all functions and classes are provided on the package website (\url{https://hesim-dev.github.io/hesim/}). 

One notable set of features not covered relates to the construction of transition probability matrices, the core of any cDTSTM. In Section~\ref{sec:cdtstm-example}, an approach was used that was tailored to a case in which transition intensity matrices were available for a reference treatment and relative treatment effects (i.e., relative risks) were available for the remaining competing interventions. To maximize efficiency, the matrix exponential was only applied where necessary (for SOC) and relative risks were applied to the transition probability matrices for SOC with the efficient \code{apply_rr()} function. 

Alternative approaches might be appropriate in other contexts. For example, if evenly spaced panel data is available, multinomial logistic regressions can be used to predict transition probabilities. On the other hand, some users may prefer a functional approach that is arguably more human readable. To facilitate this type of analysis, a \code{define_model()} block can be used to define transition probability matrices in terms of expressions that are evaluated later. Each possible combination of the parameters and input data (treatment strategies and patients) is automatically constructed and the expressions are used to create the transition probability matrices as a function of the parameters and input data. The same \code{define_model()} block may also be used to construct utility and cost models, if desired. Examples that leverage multinomial logistic regression and \code{define_model()} are provided on the package website. The latter is used to reproduce two examples (a simple Markov cohort model of an HIV treatment and a time-inhomogeneous Markov cohort model of total hip replacement) from the reference textbook by \citet{briggs2006decision}.

In Section~\ref{sec:ictstm-example}, we fit a clock-reset multi-state model and simulated outcomes with an individual-level simulation. An iCTSTM can, however, also be developed without fitting a statistical model using \proglang{R} by manually inputting the arguments (e.g., coefficients, probability distributions) of \code{params_surv} objects for each possible transition. If IPD is available for the reference treatment, \code{create_params()} can be use to obtain the baseline \code{params_surv} object from a fitted multi-state model, and relative treatment effects (e.g., hazard ratios) could be added as additional columns to the coefficients stored in the \code{params_surv} object. A similar approach was used in the partitioned survival analysis example above (Section~\ref{sec:psm-example}). A re-analysis of the total hip replacement model from the \citet{briggs2006decision} textbook using an individual-level simulation is provided on the package website to illustrate. 

A third feature worth mentioning is that a PSM can be extended to $N$ health states. The 3-state oncology case might, for instance, be extended to account for adverse events or multiple lines of therapy. One related application is the quality-adjusted time without symptoms or toxicity (Q-TWiST) method \citep{goldhirsch1989costs, lenderking1994eval}, which partitions patients into time spent (i) having toxic side effects, (ii) without symptoms or toxicities, and (iii) after progression. An example four-state model is provided on the package website. 

As mentioned in the introduction, \pkg{heemod} is the only other general purpose \proglang{R} package for cost-effectiveness modeling. It differs from \pkg{hesim} in that it focuses solely on cDTSTMs (i.e., Markov cohort models) and is not designed to facilitate integration of economic models with the statistical models used for parameter estimation. Another important difference is that \pkg{hesim} is considerably faster since all simulations are vectorized (i.e, performed at the \proglang{C++} level). Performance comparisons are provided on the package website. The time-inhomogeneous total hip replacement example from \citet{briggs2006decision} was run with both packages using 1,000 PSA iterations. At the time of writing, a cohort model in \pkg{heemod} completed in approximately 1.9 minutes while the same cohort model took approximately 1 second to run in \pkg{hesim} and an equivalent individual-level simulation in \pkg{hesim} ran in 7.4 seconds. 

There are a number of multi-state modeling \proglang{R} packages that can both fit models and make predictions. \pkg{hesim} leverages some of these for parameters estimation: \pkg{flexsurv} is used to fit parametric and flexible parametric survival and multi-state models when exact event times are available while \pkg{msm} is used to fit multi-state models when only panel data is available. A key difference is that \pkg{hesim} is designed specifically for health economic modeling whereas the other packages are more general purpose statistical packages. 

\pkg{mstate} is another package for multi-state modeling that is designed for estimation of non-parametric and semi-parametric (i.e., Cox proportional hazards) models \citep{de2011mstate}. It supports prediction of state probabilities and simulation of multi-state trajectories (to facilitate prediction for clock-reset models) from cumulative hazard functions evaluated at specific times. \citet{williams2017cost} has adapted \pkg{mstate} for cost-effectiveness modeling applications. There are however, some disadvantages of using \pkg{mstate} for health economic modeling since it was not designed for it. \code{mssample()}---the core function for simulating multi-state models---does not natively support heterogeneous target populations, PSA, simultaneous simulation of multiple treatment strategies, or computation of QALYs and costs from the simulated output. It is also less computationally efficient than \pkg{hesim} for two main reasons. First, simulation of general cumulative hazards functions is considerably slower than simulation of parametric distributions when efficient random number generators are available. Second, it is written completely in \proglang{R} and incorporating PSA, multiple treatment strategies, or heterogeneous patients results in code that is not vectorized. To compare computational performance, we ran 1,000 PSA iterations where 1 treatment strategy and 1,000 homogeneous patients were simulated through a 6-state Weibull model during each iteration. The model completed in 2.8 seconds in \pkg{hesim} and 3.1 hours in \pkg{mstate}, showing that computational considerations are not just an academic concern. 

A number of extensions of the current version of \pkg{hesim} would be helpful. First, we plan to extend the individual-level simulation so that users can update input data during the simulation as a function of time elapsed and/or the health states transitioned to. This would allow potentially time-varying variables such as age to update each time a patient transitions to a new state. Second, we would like to add more examples to ensure that \pkg{hesim} is fit for purpose for a wide range of applications in health economics. Third, more complex models of utility \citep{kharroubi2007modelling, hernandez2013relationship} and costs \citep{nixon2005methods} could help better capture heterogeneity in preferences and spending. Fourth, it would be useful to add functionality that automatically creates tunnel states for users in cDTSTMs. Fifth, equity considerations could be incorporated by adding support for distributional CEA \citep{asaria2015distributional, asaria2016distributional, cookson2017using}. Finally, a larger aspiration is to provide support for parameterization via NMA, particularly for estimation of relative treatment effects in survival (e.g., for PSMs) and multi-state models. However, since existing \proglang{R} packages are lacking in this area, this could involve creation of a separate NMA package and a new dependency for \pkg{hesim}. 

\section{Conclusion} \label{sec:conclusion}
\pkg{hesim} is a modular and computationally efficient \proglang{R} package for development and analysis of simulation models for economic evaluation. Discrete time cohort and continuous time individual-level models are supported, encompassing Markov (time-homogeneous and time-inhomogeneous) and semi-Markov processes as well as partitioned survival analysis. To facilitate integration of statistical methods and economic models and use of both patient- and aggregate-level data, models can be parameterized by either fitting statistical models using \proglang{R}, inputting values directly, or from a combination of the two. Simulated costs and QALYs from a PSA can be used for decision-analysis within a cost-effectiveness framework, allowing users to represent decision uncertainty (e.g., cost-effectiveness planes, cost-effectiveness acceptability curves, and cost-effectiveness acceptability frontiers), and conduct value of information analysis. Algorithms are efficiently coded and written in \proglang{C++} so that approaches that have been considered computationally intensive such as individual-level simulations and PSA can run very quickly. The modular design makes modeling more flexible and it easier to add new features.


\bibliography{references}



\end{document}